\newif\ifconfver
\confverfalse      
\confvertrue        

\ifconfver
     \documentclass[10pt,twocolumn,twoside]{IEEEtran}
\else
    \documentclass[11pt,draftcls,onecolumn]{IEEEtran}
\fi

\usepackage{soul}
\usepackage{graphicx}
\usepackage{psfrag}
\usepackage{url}
\usepackage{stfloats}
\usepackage{amsfonts,amssymb,amsmath,bm,paralist,theorem,cite,ifthen,color}
\usepackage{array}
\usepackage{caption}
\usepackage{calc}
\usepackage{subfigure}
\usepackage{cases}
\usepackage{verbatim}
\usepackage{algorithm,algpseudocode}
\usepackage{multirow}
\usepackage{mathrsfs}
\usepackage{epstopdf}

\DeclareMathOperator*{\argmax}{argmax}
{
\theorembodyfont{\rmfamily}
\theoremheaderfont{\itshape}

}

\newtheorem{Lemma}{\bf Lemma}
\newtheorem{Prop}{\bf Proposition}
\newtheorem{Theorem}{\bf Theorem}

\definecolor{orange}{RGB}{255,107,0}

\newlength{\twidth}
\ifconfver
\else
\fi

    \makeatletter
    \def\multilimits@{\bgroup
  \Let@
  \restore@math@cr
  \default@tag
 \baselineskip\fontdimen10 \scriptfont\tw@
 \advance\baselineskip\fontdimen12 \scriptfont\tw@
 \lineskip\thr@@\fontdimen8 \scriptfont\thr@@
 \lineskiplimit\lineskip
 \vbox\bgroup\ialign\bgroup\hfil$\m@th\scriptstyle{##}$\hfil\crcr}
    \def\Sb{_\multilimits@}
    \def\endSb{\crcr\egroup\egroup\egroup}
\makeatother

{\begin{list}{}%
    {\setlength{\rightmargin}{0pc}%
    \setlength{\leftmargin}{1pc} }} %
{\end{list}}

{\begin{list}{}%
    {\setlength{\rightmargin}{1pc}%
    \setlength{\labelsep}{0pc}
    \setlength{\leftmargin}{1pc} }} %
{\end{list}}

{\begin{list}{}%
    {\setlength{\rightmargin}{0pc}%
    \setlength{\leftmargin}{0pc} }} %
{\end{list}}

{\begin{list}{}{
    \settowidth{\labelwidth}{\mbox{\textnormal{#1}}}%
    \setlength{\leftmargin}{\labelwidth+\labelsep}}}%
{\end{list}}

\begin{document}

\bibliographystyle{IEEEtran}

\title{Sum Secrecy Rate Maximization for Full-Duplex Two-Way  Relay Networks Using Alamouti-based Rank-Two Beamforming}
\ifconfver \else {\linespread{1.1} \rm \fi

\author{
Qiang Li,~\IEEEmembership{Member,~IEEE,}  Wing-Kin Ma~\IEEEmembership{Senior~Member,~IEEE,} and Dong Han
\thanks{Qiang Li is with
School of Communication and Information Engineering, University of Electronic Science and Technology of China, Chengdu, P.~R.~China. E-mail: lq@uestc.edu.cn}
\thanks{Wing-Kin Ma is with the
Department of Electronic Engineering, The Chinese University of Hong
Kong, Shatin, Hong Kong S.A.R., China. E-mail: wkma@ieee.org.}
\thanks{Dong Han is with the Department of Electrical Engineering, The University of Texas at Dallas (UTD), Richardson, TX. E-mail: dxh151130@utdallas.edu}
}


\maketitle

\ifconfver \else
\begin{center} \vspace*{-2\baselineskip}
\end{center}
\fi
\begin{abstract}
Consider a two-way communication scenario
where two single-antenna nodes, operating under full-duplex mode, exchange information to one another through the aid of a (full-duplex) multi-antenna relay,
and there is another single-antenna node who intends to eavesdrop.
The relay employs artificial noise (AN) to interfere the eavesdropper's channel,
and amplify-forward (AF) Alamouti-based rank-two beamforming to establish the two-way communication links of the legitimate nodes.
Our problem is to optimize the rank-two beamformer and AN covariance for sum secrecy rate maximization (SSRM).
This SSRM problem is nonconvex,
and we develop an efficient solution approach using semidefinite relaxation (SDR) and minorization-maximization (MM).
We prove that SDR is tight for the SSRM problem and thus introduces no loss.
Also, we consider an inexact MM method where an approximately but computationally cheap MM solution update is used in place of the exact update in conventional MM.
We show that this inexact MM method guarantees convergence to a stationary solution to the SSRM problem.
The effectiveness of our proposed approach is further demonstrated by an energy-harvesting scenario extension, and by extensive simulation results.
\\\\
\ifconfver
\else
\noindent {\bfseries EDICS}: MSP-CODR (MIMO precoder/decoder design), MSP-APPL (Applications of MIMO communications and signal processing), SAM-BEAM (Applications of sensor and array multichannel processing)
\fi
\end{abstract}

\ifconfver \else \IEEEpeerreviewmaketitle} \fi

\ifconfver \else
\newpage
\fi


\section{Introduction} \label{sec:introduction}



With the recent advances of self-interference   cancellation   techniques, full-duplex (FD) communication has received renewed interest.
Particularly, FD has been seen as a promising physical-layer technology to meet the explosive data requirement for the future 5G mobile networks~\cite{Kim_FD}. In contrast to frequency-division duplex (FDD) and time-division duplex (TDD), FD has the potential to double the spectral efficiency by simultaneously transmitting and receiving (STR) over the same radio-frequency (RF) bands.
FD also provides new opportunities for system designs to achieve some specific goals, such as physical-layer (PHY) security~\cite{ZhengG13,lxLi,FengR,WangYP,GaoFF,Ng,Tho,ChenG,LeeJH} and simultaneous  wireless information and power transfer (SWIPT)~\cite{KangX,JuH,ZhongC,ZengY}.

PHY security is an information theoretic approach for providing information security at the PHY layer by exploiting the difference between the decoding abilities of the target user and eavesdropper.
An effective way to deliver PHY security is to adopt the so-called artificial noise (AN) approach,
where the transmitter intentionally generates noise to jam the eavesdropper.
Interestingly, with FD STR transceivers, we can apply AN even more effectively.
In \cite{ZhengG13}, the authors exploit the full duplexity of the target user to  simultaneously receive information and transmit AN.
Motivated by the aforementioned work, PHY security using FD has received considerable attention.
Specifically, the work~\cite{lxLi} analyzed the secure degrees of freedom in FD point-to-point transmission. For FD two-way secure communications, robust and low-complexity transmit solutions have been developed in~\cite{FengR} and~\cite{WangYP}, respectively. In~\cite{GaoFF} and~\cite{Ng}, the authors considered secrecy designs in cellular networks with an FD base station (BS) and multiple half-duplex uplink/downlink mobile users. For both of these works, the semidefinite relaxation-based approach was employed either to maximize the downlink secrecy rate~\cite{GaoFF} or to minimize the uplink/downlink transmission powers under secrecy rate constraints~\cite{Ng}.
FD relay secure communication has also gained much interest~\cite{Tho,ChenG,LeeJH}. In~\cite{Tho}, the authors considered  one-way FD  secure relay   networks, where two operation modes of the FD relay are considered, namely full-duplex transmission (FDT) and full-duplex jamming (FDJ).
A secrecy outage probability comparison between FDT and FDJ was conducted
in~\cite{ChenG}, and
the
result
reveals that FDJ is more suitable for the small target secrecy rate regime. Extension to a multi-hop FD relay network has also been considered in~\cite{LeeJH}.

Apart from PHY security, another emerging application of full duplexity is
SWIPT.
SWIPT is a means of using RF signals to achieve dual transmission of information and energy; readers are referred to the recent magazine paper~\cite{Krikidis} for a more complete treatment of this kind of technique. With the FD capability, a communication node can simultaneously receive information from and  broadcast energy to other nodes, or do the opposite. Under this new information-energy paradigm, various resource allocations and protocol designs have been proposed. For example, the works~\cite{KangX,JuH} considered the resource allocation problem for an information-energy hybrid cellular network, where an FD BS broadcasts energy to power users in the downlink, and at the same time each user transmits information in the uplink in a TDD manner. Optimal time and power allocations were derived in~\cite{KangX,JuH}. In~\cite{ZhongC,ZengY},  a two-hop  relay system with an FD energy harvesting relay was studied. In particular, the work~\cite{ZhongC} proposed to split the transmission into two stages for power transfer and information forwarding, and  analyzed the rate performance of FD relaying under different harvesting antenna settings and transmission modes. In~\cite{ZengY}, the authors considered a different relaying protocol to allow the relay to harvest energy and forward information concurrently.

In this work, we consider exploiting full duplexity to enhance PHY security and achieve SWIPT. Specifically, we focus on FD two-way relay networks, where two legitimate nodes simultaneously transmit and receive confidential information through an FD multi-antenna relay, and an eavesdropper overhears the transmission from both the two legitimate nodes and the relay. Unlike the previous works on FD two/multi-hop relay network~\cite{Lee15,Tho15,ChenG}, where the relay operates under either FDT mode or FDJ mode, we consider a more general relaying strategy---simultaneously relaying information and sending jamming signal. In particular, an AN-aided Alamouti-based rank-two beamforming strategy~\cite{Wu} is employed to forward the confidential information.
Our goal is to jointly optimize the rank-two beamforming matrix and the AN covariance matrix, such that the sum secrecy rate of the two-way transmission is maximized. This sum secrecy rate maximization (SSRM) problem is nonconvex by nature, but can be converted into a form suitable for minorization-maximization (MM) after applying the rank-two semidefinite relaxation (SDR) technique.
Thus, the classical MM approach~\cite{MM} can be invoked to iteratively compute a stationary solution to the relaxed SSRM problem. Since the stationary guarantee holds for the relaxed SSRM problem, but not directly for the original SSRM problem, we further develop a specific way to retrieve a stationary solution to the latter from any stationary solution to the former. The key
idea
is to exploit the rank-two beamforming structure to pin down the SDR tightness. We should point out that the classical MM approach requires solving each MM subproblem to optimality, which could be computational demanding in practice. In light of this drawback, we further propose an {\it inexact} MM approach to the SSRM problem, under which an approximate solution is sought at each iteration via an iteration-limited projected gradient method.  We prove that  the proposed  inexact MM has the same stationary convergence guarantee as the classical (exact) MM.

As an extension, we further consider the above two-way FD relay secrecy design with a wireless energy-harvesting eavesdropper. In particular, we assume that the eavesdropper is also a system user who aims to harvest energy, but could potentially eavesdrop the confidential information. In such a case, the AN plays a dual role --- on one hand, it jams the eavesdropper to secure the two-way communication; on the other hand, it also provides a source of energy for the eavesdropper to harvest. Our goal here is again to maximize the system's sum secrecy rate with the energy harvesting constraint on the eavesdropper. Following our SDR-based MM approach, we show that a stationary solution to the SSRM problem with energy harvesting can also be iteratively computed via either exact or inexact MM updates.

Our main contributions are summarized below:
\begin{itemize}
  \item  We studied a joint Alamouti-based rank-two beamforming and AN design for secrecy sum rate maximization in a full-duplex two-way relay network, with direct links between the legitimate nodes and the eavesdropper. This formulation was not considered in the prior literature.


  \item We developed an SDR-based MM approach for
  the aforementioned design formulation.
This proposed approach guarantees convergence to a stationary solution, and the proof technique, which connects SDR  tightness and MM,
 is new.


  \item Further, we proposed an inexact alternative to the SDR-based MM approach for low-complexity implementation.
   We showed that
  this inexact MM  guarantees convergence to a stationary solution.


  \item We considered a scenario extension where the eavesdropper is also an energy-harvesting user.

\end{itemize}

\subsection{Organization and Notations}
This paper is organized as follows. The system model and problem statement are given in Section~\ref{sec:system}.
Section~\ref{sec:exact_dc} focuses on the SSRM problem and develops an SDR-based MM approach.
Section~\ref{sec:in-dc} proposes an inexact MM approach to the SSRM problem. Extension
to the
energy-harvesting eavesdropper case is considered in Section~\ref{sec:ssrm-eh}.
Simulation results comparing the proposed designs are illustrated in Section~\ref{sec:simulation}. Section~\ref{sec:conclusion} concludes the paper.

Our notations are as follows.
$(\cdot)^T$, $(\cdot)^*$ and $(\cdot)^H$ denote transpose, conjugate and conjugate transpose, respectively;  $\mathbf{I}$ denotes an identity matrix with appropriate dimension; $\mathbb{H}_{+}^{N}$ denotes the set of all $N$-by-$N$ Hermitian positive semidefinite matrices; $\mathbf{A}\succeq \mathbf{0}$ means that $\mathbf{A}$ is Hermitian positive semidefinite, and ${\bf A}\succ {\bf 0}$ means that $\mathbf{A}$ is Hermitian positive definite;
${\rm Diag}(a, ~b)$ represents a diagonal matrix with diagonal elements $a$ and $b$;
$[ \cdot ]^+$ is the projection onto the set of non-negative numbers; $\mathcal{CN}(\bm a, \bm \Sigma)$ represents complex Gaussian distribution with mean $\bm a$ and covariance matrix $\bm \Sigma$.

\section{System Model and Problem Formulation}\label{sec:system}

\begin{figure}[htp!]
\centerline{\resizebox{.4\textwidth}{!}{\includegraphics{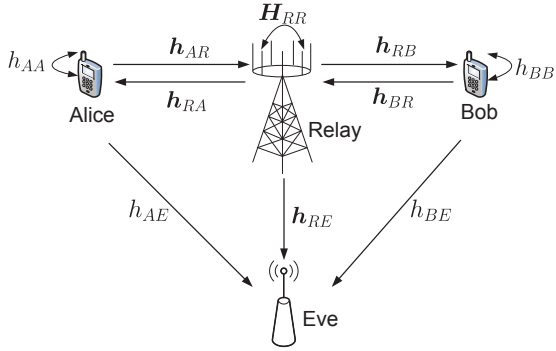}}
}
\vspace{-.5\baselineskip}
\caption{System model.} \label{fig:model}
\end{figure}

The scenario of interest is depicted in Fig.~\ref{fig:model}.
Two legitimate nodes, named Alice and Bob herein, perform two-way communication with the aid of a relay.
Alice, Bob and the relay are equipped with full-duplex RF transceivers,
and thus they can simultaneously transmit and receive over the same RF band.
Alice has one antenna for transmission and one antenna for reception,
and the same applies to Bob.
The relay has $N$ antennas for transmission and $M$ antennas for reception.
The transmission is overheard by an evesdropper, named Eve, who has one antenna.
The problem is to design a transmission scheme such that the two-way messages are both secured from a PHY information security perspective.

\subsection{Received Signal Model at Alice and Bob}

Let us first describe the basic signal model.
Alice and Bob transmit coded confidential information signals $s_A(t) \in \mathbb{C}$ and $s_B(t) \in \mathbb{C}$ to Bob and Alice, respectively (resp.).
There is no direct link between Alice and Bob,
and the information signals are forwarded by the relay.
Let $\bm h_{i,R} \in \mathbb{C}^M$, $i\in \{A, B\}$, denote the channel from node $i$ to the receive antennas of the  relay,
where nodes $A$ and $B$ refer to Alice and Bob, resp.
Also, let $\bm H_{RR} \in \mathbb{C}^{M\times N}$ denote the channel from the transmit antennas to the receive antennas of the relay, i.e., the so-called self-interference (SI) channel.
The relay's received signal at   the $t$th time index  is modeled as
\begin{equation} \label{eq:rx_sig_relay}
\begin{aligned}
  \bm y_R(t) = \bm h_{AR} s_A(t)  + \bm h_{BR} s_B(t) +  \bm H_{RR} & \bm x_R(t) +  \bm n_R(t), \\
  &  ~~t=1,2,\ldots
   \end{aligned}
\end{equation}
where
$\bm x_R(t) \in \mathbb{C}^{N}$ is the transmit  signal at the relay, which aims at simultaneously forwarding the confidential information signals to Bob and Alice;
 $\bm n_R(t)\sim\mathcal{CN}(\bm 0, \sigma_R^2 \bm I)$ is additive white Gaussian noise (AWGN).
 The format that $\bm x_R(t)$ takes will be described soon.
 Note that the third term $\bm H_{RR} \bm x_R(t)$ on the right-hand side (RHS) of \eqref{eq:rx_sig_relay} is the full-duplex SI, which has a much larger dynamic range than the signal components $\bm h_{AR} s_A(t)$ and $\bm h_{BR} s_B(t)$.
 However, the full-duplex SI
 can be eliminated  by applying nulling, where $\bm x_R(t)$ is designed such that $\bm H_{RR} \bm x_R(t)= \bm 0$~\cite{ZhengG}.
 We will take on this SI nulling strategy\footnote{Implicitly, we have assumed $N>M$ so that nulling can be done at the relay's transmission stage. On the other hand,  for the case of $N<M$, a similar nulling process can be performed at the relay's reception stage; readers are referred to our previous abridged conference paper~\cite{qli_icassp16} for the latter case.}.
 Moreover, the received signals at Alice and Bob are
\begin{equation} \label{eq:rx_sig_nodei}
 \begin{aligned}
   y_i(t) = &  \bm h_{R,i}^H  \bm x_R(t) + h_{i,i} s_i(t) + n_i(t), \\
& \qquad \qquad \qquad \qquad ~i \in \{A,B \}, ~t=1,2,\ldots
 \end{aligned}
 \end{equation}
 where, for $i \in \{A, B \}$,
 $y_i(t)$ is the received signal at node $i$;
 $\bm h_{R,i} \in \mathbb{C}^N$ is the channel from the relay to node $i$;
 $n_i(t) \sim \mathcal{CN}( 0, \sigma_i^2 )$ is AWGN.
 Again, the second term $h_{i,i} s_i(t)$ on the RHS of \eqref{eq:rx_sig_nodei} is full-duplex SI.
 Since Alice and Bob have one receive antenna only, the aforementioned SI nulling strategy is not applicable.
 However, since Alice (resp. Bob) has knowledge of its own transmitted signal $s_A(t)$ (resp. $s_B(t)$),  it can cancel the SI term $h_{AA} s_A(t)$ (resp. $h_{BB} s_B(t)$) from \eqref{eq:rx_sig_nodei}.
Such signal cancellation is however imperfect owing to issues such as the large dynamic range of the full-duplex SI; see~\cite{Kim_FD} for details.
The SI-cancelled signal of $y_i(t)$ can be modeled as
\[
\begin{aligned}
\bar{y}_i(t) = \bm h_{R,i}^H  \bm x_R(t) +  \sqrt{\kappa_i} h_{i,i} & s_i(t) + n_i(t), \\
&~~ i \in \{A,B \}, ~ t=1,2,\ldots
\end{aligned}
\]
where $0 < \kappa_i < 1$ is the full-duplex SI residual factor of node $i$~\cite{ZhongC}.

The transmission format of $\bm x_{R}(t)$ is specified as follows.
We consider a combination of amplify-and-forward (AF) beamforming, Alamouti space-time coding and artificial noise (AN) strategies.
As mentioned previously, we assume SI nulling where  the received signal in \eqref{eq:rx_sig_relay} is reduced to
$ \bm y_R(t) = \bm h_{AR} s_A(t)  + \bm h_{BR} s_B(t)  + \bm n_R(t)$.
The relay first obtains soft estimates of $s_A(t)$ and $s_B(t)$ via the minimum mean-square-error (MMSE) reception
\begin{equation*}
        \hat{s}_i(t)  = \bm f_i^H  \bm y_R(t), ~~i\in \{A, B\},
    \end{equation*}
where
$\bm f_A = (\sigma_R^2 \bm I + \sum_{i\in \{A, B\}}  p_i \bm h_{iR} \bm h_{iR}^H )^{-1} \bm h_{AR}$; $\bm f_B = (\sigma_R^2 \bm I + \sum_{i\in \{A, B\}}  p_i \bm h_{iR} \bm h_{iR}^H )^{-1} \bm h_{BR}$; $p_i= \mathbb{E}\{ |s_i(t)|^2 \} $, $i \in \{ A, B \}$.
Then, the estimates $\hat{s}_A(t)$ and $\hat{s}_B(t)$ are parsed into blocks via
$\hat{\bm s}_i(n)= [ \hat{s}_i(2n-1)~ \hat{s}_i(2n)]^T$,
where $i \in \{ A, B \}$, and $n=1,2,\ldots$  denotes the block index.
Similarly, let $\bm X_R(n) = [ \bm x_R(2n-1) ~ \bm x_R(2n) ]$ denote the $n$th block of the relay's transmit signal.
At every block $n$, the relay uses an AF-beamformed Alamouti scheme to forward the  estimated information blocks $\hat{\bm s}_A(n)$ and $\hat{\bm s}_B(n)$.
To be specific, we have
\begin{equation}\label{eq:tx_sig_relay}
  {\bm X}_R(n) = \bm W_0 \bm C\left(\hat{\bm s}_A(n  ) + \hat{\bm s}_B(n)\right) + \bm Z(n),
\end{equation}
where $\bm W_0 \in \mathbb{C}^{N \times 2}$ is  a transmit beamforming matrix; \[\bm C(\bm s) = \begin{bmatrix}
  s_1  & s_2^* \\ s_2 & -s_1^*
\end{bmatrix}\]
is the Alamouti space-time code;
$\bm Z(n) = [ \bm z(2n-1) ~ \bm z(2n) ] \in \mathbb{C}^{N \times 2}$ is a superimposed AN for jamming Eve, 
in which $\bm z(t)$
follows an i.i.d. complex Gaussian distribution with mean zero and covariance $\bm Q_0 \in \mathbb{H}_+^N$.
To fulfill the SI nulling condition $\bm H_{RR} \bm x_R(n) = \bm 0$,
$\bm W_0$ and $\bm Q_0$ are restricted to satisfy $\bm H_{RR} \bm W_0 = \bm 0$ and $\bm H_{RR} \bm Q_0 = \bm 0$, resp.
Let us examine the corresponding received signals at Alice and Bob.
By letting $\bar{\bm y}_i(n) = [ \bar{y}_i(2n-1) ~ \bar{y}_i(2n) ]$,
$\bm s_i(n)= [ s_i(2n-1) ~ s_i(2n) ]$,
$i \in \{ A, B \}$,
we have
\begin{equation}\label{eq:rx_sig}
\begin{aligned}
 \bar{\bm y}_i(n)   = & \bm h_{R,i}^H \bm W_0 \bm C\left(\hat{\bm s}_B(n )\right) +
\bm h_{R,i}^H \bm W_0 \bm C\left(\hat{\bm s}_A(n ) \right)
\\
& + \bm h_{R,i}^H \bm Z(n) +
\sqrt{\kappa_i} h_{i,i}   \bm s_i(n+1)
+ \bm n_i(n).
\end{aligned}
\end{equation}
Suppose $i = A$. Then,
the terms
involving $s_A(2n-1)$ and $s_A(2n)$ in the above equation is the AF-induced self-interference,
and it can be eliminated by direct cancellation~\cite{Cui}.
Subsequently, by applying the standard Alamouti reception~\cite{Alamouti} to the SI-cancelled version of $\bar{\bm y}_A(n)$,
it can be verified that
the elements of $\bm s_B(n)$
can be decoupled from $\bar{\bm y}_A(n)$ with the same signal-to-interference-pluse-noise ratio (SINR).
Particularly, it can be shown from the above system setup that the SINR of $s_B(n)$ at Alice is
\[{\sf SINR}_A (\bm W_0, \bm Q_0) = \frac{\tilde{p}_B  \| \bm h_{RA}^H \bm W_0\|^2 }{\tilde{\sigma}_R^2 \| \bm h_{RA}^H \bm W_0 \|^2 + \bm h_{RA}^H \bm Q_0 \bm h_{RA} + \tilde{\sigma}_A^2}, \]
where $\tilde{p}_B \triangleq  p_B |(\bm f_A + \bm f_B)^H \bm h_{BR}|^2$, $\tilde{\sigma}_R^2 \triangleq \sigma_R^2\| \bm f_A + \bm f_B \|^2$, $\tilde{\sigma}_A^2 \triangleq \sigma_A^2 + \kappa_A p_A |h_{AA}|^2$.
Also, the above argument applies to $i=B$,
and the SINR of $s_A(t)$ at Bob is
\begin{equation*}
  {\sf SINR}_B(\bm W_0, \bm Q_0) = \frac{\tilde{p}_A \| \bm h_{RB}^H \bm W_0\|^2 }{\tilde{\sigma}_R^2 \| \bm h_{RB}^H \bm W_0 \|^2 + \bm h_{RB}^H \bm Q_0 \bm h_{RB} + \tilde{\sigma}_B^2},
\end{equation*}
where $\tilde{p}_A  \triangleq p_A |(\bm f_A + \bm f_B)^H \bm h_{AR}|^2 $, $\tilde{\sigma}_B^2 \triangleq \sigma_B^2 + \kappa_B p_B |h_{BB}|^2$.

\subsection{Received Signal Model at Eve}

Let us consider the received signal model at Eve under the aforementioned system setup.
Eve's received signal is modeled as
\begin{equation} \label{eq:yEt}
\begin{aligned}
y_E(t)= \bm h_{RE}^H \bm x_R(t) + h_{AE} s_A(t) + & h_{BE} s_B(t) + n_E(t),  \\
& ~~t=1,2.\ldots
\end{aligned}
\end{equation}
where
$\bm h_{RE} \in \mathbb{C}^N$ is the channel from the relay to Eve;
$h_{i,E} \in \mathbb{C}$, $i \in \{ A, B \}$, is the channel from node $i$ to Eve;
$n_E(t) \sim \mathcal{CN}(0, \sigma_E^2)$ is noise.
Note that in the above model, there are direct links between the legitimate nodes and Eve.
Denote $\bm y_E(n) = [ y_E(2n-1) ~ y_E(2n) ]$.
Similar to \eqref{eq:rx_sig}, it can be shown that
\begin{equation} \label{eq:yEn}
\begin{aligned}
 \bm y_E(n)   = & \bm h_{RE}^H \bm W_0 \bm C\left(\hat{\bm s}_B(n )\right) +
\bm h_{RE}^H \bm W_0 \bm C\left(\hat{\bm s}_A(n ) \right)
\\
& + \bm h_{RE}^H \bm Z(n) +
{\textstyle \sum_{i \in \{ A, B \} } h_{i,E} \bm s_i(n+1) }
+ \bm n_E(n).
\end{aligned}
\end{equation}
From the above formula, we observe that $\bm s_A(n)$ and $\bm s_B(n)$ are present in both $\bm y_E(n)$ and $\bm y_E(n-1)$.
Let us assume that Eve intends to decode $\bm s_A(n)$ and $\bm s_B(n)$ from $\bm y_E(n)$ and $\bm y_E(n-1)$, seeing other terms as interference and noise.
Then, through some tedious derivations which are shown in Appendix \ref{sec:Appendix-Eve-model},
Eve's reception can be equivalently expressed as an MIMO system
\begin{equation}\label{eq:mimo_channel_Eve}
  \tilde{\bm y}_E(t) ={\bm H}_E
  \begin{bmatrix}
    s_A(t)\\
    s_B(t)
  \end{bmatrix}  + \tilde{\bm n}_E(t) \in \mathbb{C}^2, ~~t=1,2,\ldots.
\end{equation}
Here,
$$  \bm H_E =  \begin{bmatrix}
    \| \bm h_{RE}^H \bm W_0\|   \tilde{f}_{AR} & \| \bm h_{RE}^H \bm W_0\|   \tilde{f}_{BR} \\
       h_{AE} &    h_{BE}
  \end{bmatrix}$$
 is the equivalent MIMO channel, where
  $\tilde{f}_{iR} = (\bm f_A + \bm f_B)^H \bm h_{iR}, i\in \{A, B\}$;
  $\tilde{\bm n}_E(t)$ is the interference-plus-noise term with mean zero and covariance
   \begin{equation}\label{eq:def_Psi}
  \bm \Psi = \begin{bmatrix}
    \Psi_{11} &  0 \\
     0 &   \Psi_{22}
  \end{bmatrix} \in \mathbb{R}^{2\times 2},
  \end{equation}
where $\Psi_{11} = \tilde{\sigma}_R^2  \| \bm h_{RE}^H \bm W_0\|^2 + \bm h_{RE}^H \bm Q_0 \bm h_{RE} + \sigma_E^2 + \sum_{i\in\{A,B\}} p_i|h_{iE}|^2$ and $\Psi_{22} =\bm h_{RE}^H \bm Q_0 \bm h_{RE} + \sigma_E^2 +  (\tilde{p}_A+\tilde{p}_B+\tilde{\sigma}_R^2)\| \bm h_{RE}^H \bm W_0\|^2$.

\subsection{Sum Secrecy  Rate Maximization Problem}

Under the system setup in the last two subsections, the problem is to design the AF beamforming matrix $\bm W_0$ and the AN covariance $\bm Q_0$ such that the sum secrecy rate of Alice and Bob is maximized under a total transmission power constraint at the relay.
The secrecy achievable rate formulation is as follows.
The achievable rates at Alice and Bob are modeled as
\begin{equation}\label{eq:rate_AB}
  R_i(\bm W_0, \bm Q_0) = \log \left( 1 + {\sf SINR}_i(\bm W_0, \bm Q_0) \right), ~ i \in \{ A, B \},
\end{equation}
where we treat the full-duplex residual SI as Gaussian noise; see~\cite{ZhongC,ZhengG} for similar treatments.
Also, by applying the MIMO
 multiple-access channel capacity result to \eqref{eq:mimo_channel_Eve}, the sum achievable rate at Eve is formulated as
\begin{equation} \label{eq:rate_Eve_0}
 R_{E} (\bm W_0, \bm Q_0) = \log | \bm I + \bm H_E \bm P \bm H_E^H \bm \Psi^{-1} |,
\end{equation}
where $\bm P = {\rm Diag}(p_A, ~p_B)$.
The above achievable rate can be reduced to
\begin{equation}\label{eq:rate_Eve}
\begin{aligned}
  & R_{E} (\bm W_0, \bm Q_0) \\
   = & \log \left( \frac{\Psi_{22}^2 + \theta_1(\Psi_{22}+\Psi_{11}) +  \theta_2 \| \bm h_{RE}^H \bm W_0\|^2 }{\Psi_{11} \Psi_{22}} \right)
\end{aligned}
\end{equation}
where
$\theta_1  =  \textstyle \sum_{i\in \{A, B\}} p_i |h_{iE}|^2$,
$\theta_2  =   (\textstyle \sum_{ i\in \{A, B\} } p_i |\tilde{f}_{iR}|^2)
\theta_1
$  $- |\textstyle \sum_{i\in \{A, B\}} \tilde{f}_{iR}^* h_{iE}|^2$;
see Appendix~\ref{sec:Appendix-Eve-model-2} for details.
From~\eqref{eq:rate_Eve} and \eqref{eq:rate_AB}, we characterize the  sum secrecy rate as~\cite{Yener}\footnote{The sum secrecy rate $R_s$ implicitly assumes that Alice and Bob can coordinately allocate their transmission rates.
This can be made possible by
asking the relay to coordinate the rate allocation.
}
\[
R_s(\bm W_0, \bm Q_0) \triangleq  R_A(\bm W_0, \bm Q_0) + R_B(\bm W_0, \bm Q_0) - R_{E}(\bm W_0, \bm Q_0) .\]
Moreover, from the system setup in the last subsection,
it can be verified that the total transmission power at the relay is
\begin{align*}
   p_R (\bm W_0, \bm Q_0)  = &  \frac{1}{2} {\rm Tr} (\mathbb{E} \{\bm X_R(n) \bm X_R(n)^H\} ) \\
   = & \zeta {\rm Tr}(\bm W_0 \bm W_0^H) +  {\rm Tr}(\bm Q_0),
 \end{align*}
where $\zeta = p_A | (\bm f_A + \bm f_B)^H \bm h_{AR}|^2 + p_B | (\bm f_A + \bm f_B)^H \bm h_{BR}|^2 + \sigma_R^2 \| \bm f_A + \bm f_B\|^2$.
The design problem is therefore formulated as follows:
\begin{subequations} \label{eq:main_prob}
  \begin{align}
     \max_{\bm Q_0 \succeq \bm 0, ~\bm W_0 \in \mathbb{C}^{N\times 2}} ~~ & R_s(\bm W_0, \bm Q_0)
     \label{eq:main_prob_a} \\
   {\rm s.t.} ~ ~&  p_R({\bm W}_0, {\bm Q}_0) \leq P_R,  \label{eq:main_prob_b}\\
    ~ ~&    \bm H_{RR} \bm W_0 = \bm 0 , ~ \bm H_{RR} \bm Q_0 = \bm 0, \label{eq:main_prob_c}
  \end{align}
\end{subequations}
where $P_R>0$ is the maximal transmission power threshold at the relay, and
recall that \eqref{eq:main_prob_c} is for fulfilling the full-duplex SI nulling condition.
Problem~\eqref{eq:main_prob} will be called the {\em sum secrecy  rate maximization} (SSRM) problem in the sequel.

\section{An SDR-based MM Approach to the SSRM Problem }\label{sec:exact_dc}

The SSRM problem in \eqref{eq:main_prob} is nonconvex,
and our aim is to develop an SDR-based  minorization-maximization (MM) approach that will be shown to guarantee convergence to a stationary solution to problem \eqref{eq:main_prob}.
In the first subsection, we give a detailed   description of our proposed approach, and
 in
the second subsection
 we show
the convergence of the SDR-based MM algorithm.


\subsection{Description of the SDR-based MM Algorithm } \label{subsec:mm_alg}
Our development is as follows. First, we consider an alternative formulation of problem~\eqref{eq:main_prob}.
  Let $r={\rm rank}(\bm H_{RR} )$ and $\bm V_0 \in \mathbb{C}^{N\times (N-r)}$ be the right singular vectors associated with the zero singular values of $\bm H_{RR}$.
 From \eqref{eq:main_prob_c}, it is easy to verify that
 any feasible point $(\bm W_0, \bm Q_0)$ of problem~\eqref{eq:main_prob} can be equivalently expressed as
\begin{equation*} \label{eq:append_a_1}
  \bm W_0 = \bm V_0 \tilde{\bm W}, ~~ \bm Q_0 = \bm V_0 \bm Q \bm V_0^H,
\end{equation*}
for some $\tilde{\bm W} \in \mathbb{C}^{(N-r) \times 2}$ and $\bm Q \in \mathbb{H}_+^{n-r}$.
By applying the above equivalence to problem~\eqref{eq:main_prob}, and letting $\tilde{\bm W} = \zeta^{-1/2} \bm W$, we can rewrite problem~\eqref{eq:main_prob} as
\begin{equation} \label{eq:main_dlinkeqv}
  \begin{aligned}
    \max_{\bm Q \in \mathbb{H}_+^{N-r},~{\bm W} \in \mathbb{C}^{(N-r) \times 2}} ~ &  \phi(\bm W \bm W^H, \bm Q)   \\
    {\rm s.t.} ~ & {\rm Tr}(\bm W \bm W^H) + {\rm Tr}(\bm Q) \leq P_R,
  \end{aligned}
\end{equation}
where $ \phi(\bm W \bm W^H, \bm Q) \triangleq f(\bm W \bm W^H, \bm Q) - g_1(\bm W \bm W^H, \bm Q)- g_2(\bm W \bm W^H, \bm Q) $, and
$f$, $g_1$ and $g_2$ are defined in~\eqref{eq:alpha_beta_def} on the top of the next page.
Also, it is immediate that  the solutions  of problems~\eqref{eq:main_prob} and \eqref{eq:main_dlinkeqv} are related through $\bm W_0= \zeta^{1/2} \bm V_0 \bm W$ and $\bm Q_0  = \bm V_0 \bm Q \bm V_0^H$.
\begin{figure*}[!t]
\setlength\arraycolsep{0pt}
 \begin{equation}\label{eq:alpha_beta_def}
  \begin{aligned}
  f (\bm W \bm W^H, \bm Q) & = \textstyle\sum_{i=1}^2 \log(c_i+ \alpha_i(\bm W \bm W^H, \bm Q)) + \log(  \psi_{1}(\bm W \bm W^H, \bm Q)) + \log(  \psi_{2}(\bm W  \bm W^H, \bm Q)), \\
   g_1(\bm W \bm W^H, \bm Q)  & = \textstyle \sum_{i=1}^2 \log(c_i+ \beta_i(\bm W \bm W^H, \bm Q)),  \\
  g_2(\bm W \bm W^H, \bm Q)  & = \log \Big(  [  \psi_{2}(\bm W \bm W^H, \bm Q)]^2 + \theta_1 \big[  \psi_{1}(\bm W \bm W^H, \bm Q)+  \psi_{2}(\bm W \bm W^H, \bm Q) \big] +  \zeta^{-1}\theta_2 {\rm Tr} \big( \bm W \bm W^H\bm V_0^H \bm h_{RE}\bm h_{RE}^H \bm V_0 \big) \Big),\\
 \alpha_1(\bm W\bm W^H, \bm Q) & =   \beta_1(\bm W\bm W^H, \bm Q) +  \zeta^{-1}\tilde{p}_B {\rm Tr}( \bm W \bm W^H \bm V_0^H \bm h_{RA}\bm h_{RA}^H \bm V_0), \\
\alpha_2 (\bm W\bm W^H, \bm Q) & =   \beta_2(\bm W\bm W^H, \bm Q) +  \zeta^{-1}\tilde{p}_A {\rm Tr}( \bm W \bm W^H \bm V_0^H \bm h_{RB}\bm h_{RB}^H \bm V_0), \\
 \beta_1(\bm W\bm W^H, \bm Q) & =   \zeta^{-1}\tilde{\sigma}_R^2 {\rm Tr}(\bm W \bm W^H \bm V_0^H \bm h_{RA}\bm h_{RA}^H \bm V_0) +  {\rm Tr}(\bm Q \bm V_0^H \bm h_{RA}\bm h_{RA}^H \bm V_0), \\
\beta_2(\bm W\bm W^H, \bm Q) & =    \zeta^{-1}\tilde{\sigma}_R^2 {\rm Tr}(\bm W \bm W^H \bm V_0^H \bm h_{RB}\bm h_{RB}^H \bm V_0) +  {\rm Tr}(\bm Q \bm V_0^H \bm h_{RB}\bm h_{RB}^H \bm V_0), \\
 \psi_{1}(\bm W \bm W^H, \bm Q) &  =   \tilde{\sigma}_R^2 \zeta^{-1} {\rm Tr}( \bm W \bm W^H \bm V_0^H \bm h_{RE}\bm h_{RE}^H \bm V_0) +  {\rm Tr}(\bm Q \bm V_0^H \bm h_{RE}\bm h_{RE}^H \bm V_0) + \sigma_E^2 +  \textstyle \sum_{i\in\{A,B\}} p_i|h_{iE}|^2, \\
  \psi_{2}(\bm W \bm W^H, \bm Q) & =   {\rm Tr}(\bm Q \bm V_0^H \bm h_{RE}\bm h_{RE}^H \bm V_0) + \sigma_E^2 +     {\rm Tr}( \bm W \bm W^H \bm V_0^H \bm h_{RE}\bm h_{RE}^H \bm V_0), \\
c_1    = \tilde{\sigma}_A^2, ~c_2  & = \tilde{\sigma}_B^2,~ \zeta = p_A | (\bm f_A + \bm f_B)^H \bm h_{AR}|^2 + p_B | (\bm f_A + \bm f_B)^H \bm h_{BR}|^2 + \sigma_R^2 \| \bm f_A + \bm f_B\|^2.
\end{aligned}
\end{equation}
     \hrulefill
    \end{figure*}

Second, we apply semidefinite relaxation (SDR) \cite{SDR} to problem~\eqref{eq:main_dlinkeqv}.
Specifically, by noticing the following equivalence
 \[ \bm{\mathcal W} = \bm W \bm W^H  \Longleftrightarrow \bm{\mathcal W} \succeq \bm 0, ~ {\rm rank}(\bm{\mathcal W}) \leq 2, \] we replace $\bm W \bm W^H$ in problem~\eqref{eq:main_dlinkeqv} with $\bm{\mathcal W}$ and drop the nonconvex rank-two constraint on $\bm{\mathcal W}$ to get an SDR of problem~\eqref{eq:main_dlinkeqv} as follows:
\begin{subequations} \label{eq:main_eqv_sdr}
  \begin{align}
    \max_{\bm{\mathcal W}\succeq \bm 0, \bm Q\succeq \bm 0} ~ & \phi(\bm{\mathcal W}, \bm Q)\triangleq f(\bm{\mathcal W}, \bm Q) - g_1(\bm{\mathcal W}, \bm Q)- g_2(\bm{\mathcal W}, \bm Q)\label{eq:main_eqv_sdr_a} \\
    {\rm s.t.} ~ & {\rm Tr}(\bm{\mathcal W}) + {\rm Tr}(\bm Q) \leq P_R.  \label{eq:main_eqv_sdr_b}
  \end{align}
\end{subequations}
Notice that $f(\bm{\mathcal W}, \bm Q)$ and $g_1(\bm{\mathcal W}, \bm Q)$ are both concave with respect to (w.r.t.) $(\bm{\mathcal W}, \bm Q)$, whereas $g_2(\bm{\mathcal W}, \bm Q)$ is neither convex nor concave w.r.t. $(\bm{\mathcal W}, \bm Q)$. Hence,
problem~\eqref{eq:main_eqv_sdr} is still a nonconvex optimization problem. In the sequel, we  develop an MM approach to~\eqref{eq:main_eqv_sdr} by finding a tight concave lower bound on $ \phi(\bm{\mathcal W}, \bm Q)$.

First of all, given any feasible point $(\hat{\bm{\mathcal{W}}}, \hat{\bm Q})$ of \eqref{eq:main_eqv_sdr}, an upper bound on $g_1(\bm{\mathcal W}, \bm Q)$ can be easily obtained from the first-order condition, or linearization of $g_1$ at $(\hat{\bm{\mathcal{W}}}, \hat{\bm Q})$, i.e.,
\begin{align*}
   & g_1(\bm{\mathcal W}, \bm Q)     \leq   \tilde{g}_1(\bm{\mathcal W}, \bm Q; \hat{\bm{\mathcal{W}}}, \hat{\bm Q}),
 \end{align*}
where $\tilde{g}_1(\bm{\mathcal W}, \bm Q; \hat{\bm{\mathcal{W}}}, \hat{\bm Q}) \triangleq  g_1(\hat{\bm{\mathcal{W}}}, \hat{\bm Q}) +   {\rm Tr}  ( \nabla_{\bm {\mathcal W}} g_1(\hat{\bm{\mathcal{W}}}, \hat{\bm Q})^H (\bm{\mathcal W} - \hat{\bm {\mathcal W}})  ) +   {\rm Tr}  ( \nabla_{\bm Q } g_1(\hat{\bm{\mathcal{W}}}, \hat{\bm Q})^H (\bm Q - \hat{\bm Q})  )$.
For $g_2(\bm{\mathcal W}, \bm Q) $, we make use of  the inequality $\log(x) \leq \log(\hat{x}) + \frac{x-\hat{x}}{\hat{x}}$ to obtain
\begin{align*}
   & g_2(\bm{\mathcal W}, \bm Q)     \leq    \tilde{g}_2(\bm{\mathcal W}, \bm Q; \hat{\bm{\mathcal{W}}}, \hat{\bm Q}),
 \end{align*}
where
\begin{align*}
  & \tilde{g}_2(\bm{\mathcal W}, \bm Q;\hat{\bm{\mathcal{W}}}, \hat{\bm Q})
  \triangleq    g_2(\hat{\bm{\mathcal{W}}}, \hat{\bm Q}) -1 + \\
     & \qquad \frac{   (\psi_{2})^2 + \theta_1(  \psi_{1}+  \psi_{2}) +  \theta_2 \zeta^{-1}  \bm h_{RE}^H \bm V_0 \bm{\mathcal W} \bm V_0^H \bm h_{RE}}{ (  \hat\psi_{2})^2 + \theta_1(  \hat\psi_{1}+  \hat\psi_{2}) +  \theta_2 \zeta^{-1}  \bm h_{RE}^H \bm V_0 \hat{\bm{\mathcal W}} \bm V_0^H \bm h_{RE} },
\end{align*} and for notational simplicity  we have dropped the arguments of $    \psi_{i}$ and used $   \hat{\psi_{i}}$ to represent $   \psi_{i}(\hat{\bm{\mathcal W}}, \hat{\bm Q})$.

Now, the proposed  MM algorithm recursively solves the following  optimization problem
\begin{subequations} \label{eq:dc_sub}
  \begin{align}
(    \bm{\mathcal W}^{k+1}, \bm Q^{k+1} ) \in  \argmax_{\bm{\mathcal W} \succeq \bm 0, \bm Q \succeq \bm 0} ~ & \tilde{\phi}(\bm{\mathcal W}, \bm Q; \bm{\mathcal{W}}^k, \bm Q^k)\label{eq:dc_sub_a} \\
    {\rm s.t.} ~ & {\rm Tr}(\bm{\mathcal W}) + {\rm Tr}(\bm Q) \leq P_R,  \label{eq:dc_sub_b}
  \end{align}
\end{subequations}
until some stopping rule is met. In~\eqref{eq:dc_sub}, we have defined  $\tilde{\phi}(\bm{\mathcal W}, \bm Q; \bm{\mathcal{W}}^k, \bm Q^k)     \triangleq f(\bm{\mathcal W}, \bm Q) - \tilde{g}_1(\bm{\mathcal W}, \bm Q; \bm{\mathcal{W}}^k, \bm Q^k) -\tilde{g}_2(\bm{\mathcal W}, \bm Q; \bm{\mathcal{W}}^k, \bm Q^k) $ for $k=0, 1, \ldots$ and some feasible starting point $(\bm{\mathcal W}^{0}, \bm Q^{0} )$.

Problem~\eqref{eq:dc_sub} is  a convex problem, which can be optimally solved, e.g., by \texttt{CVX}~\cite{cvx}.
 Moreover, by direct application of the MM convergence result~\cite[Theorem 1]{BSUM}, we
conclude that
every limit point of $\{(\bm{\mathcal{W}}^k, \bm Q^k)\}_k$ is a stationary solution to problem~\eqref{eq:main_eqv_sdr}.

The MM approach proposed above is, at first look, an MM approach to the relaxed SSRM problem in \eqref{eq:main_eqv_sdr}, rather than the original SSRM problem.
%
Thus, it is important to understand whether problem \eqref{eq:main_eqv_sdr} can provide a tight relaxation to the original SSRM problem~\eqref{eq:main_dlinkeqv},
and whether we can obtain a stationary solution to the original SSRM problem from the proposed MM iteration. We address these issues in the next subsection.

\subsection{SDR Tightness and Stationary Convergence}
In this subsection, we establish the SDR tightness and the stationary convergence of the MM iteration.
Since problem~\eqref{eq:dc_sub} has a compact  feasible set, the iterate $\{(\bm{\mathcal{W}}^k, \bm Q^k)\}_k$  generated by the MM iteration in \eqref{eq:dc_sub} has at least one limit point. Suppose that $(\bar{\bm{\mathcal{W}}}, \bar{\bm Q})$ is any limit point of  $\{(\bm{\mathcal{W}}^k, \bm Q^k)\}_k$.
Consider the following two problems:
\begin{equation}\label{eq:rank1_key_1}
  \begin{aligned}
\max_{\bm{\mathcal W} \succeq \bm 0, \bm Q \succeq \bm 0} ~ & \tilde{\phi}(\bm{\mathcal W}, \bm Q; \bar{\bm{\mathcal{W}}}, \bar{\bm Q}) \\
    {\rm s.t.} ~ & {\rm Tr}(\bm{\mathcal W}) + {\rm Tr}(\bm Q) \leq P_R,
  \end{aligned}
\end{equation}
and
\begin{subequations}\label{eq:rank1_key_3}
    \begin{align}
    &\hspace{-0.5cm} \min_{\bm{\mathcal W} \succeq \bm 0, \bm Q \succeq \bm 0} ~  {\rm Tr}(\bm {\mathcal W}) + {\rm Tr}(\bm Q)   \label{eq:rank1_key_3a}\\
    &    {\rm s.t.}~    \alpha_i(\bm{\mathcal W}, \bm Q) = \alpha_i(\bar{\bm{\mathcal{W}}}, \bar{\bm Q}), ~i=1,2, \label{eq:rank1_key_3b} \\
        &  ~~~    \beta_i(\bm{\mathcal W}, \bm Q) =  \beta_i(\bar{\bm{\mathcal{W}}}, \bar{\bm Q}), ~i=1,2, \label{eq:rank1_key_3d}\\
    &  ~~~    \psi_{i}(\bm{\mathcal W}, \bm Q) =   \psi_{i}(\bar{\bm {\mathcal W}}, \bar{\bm Q}), ~i=1,2, \label{eq:rank1_key_3c}\\
    & ~       {\rm Tr} \big( \bm{\mathcal W} \bm V_0^H \bm h_{RE}\bm h_{RE}^H \bm V_0 \big)=  {\rm Tr} \big( \bar{\bm{\mathcal W}}\bm V_0^H \bm h_{RE}\bm h_{RE}^H \bm V_0 \big), \label{eq:rank1_key_3e}
  \end{align}
\end{subequations}
where $\alpha_i$, $\beta_i$ and $\tilde{\phi}$  are defined in \eqref{eq:alpha_beta_def} and \eqref{eq:dc_sub_a}, resp. Problems~\eqref{eq:rank1_key_1} and \eqref{eq:rank1_key_3}
are closely related,
as revealed by the following lemma.
\begin{Lemma}\label{lemma:1}
Let $({\bm {\mathcal W}}^\star, {\bm Q}^\star)$  be any optimal solution to problem~\eqref{eq:rank1_key_3}. Then, $({\bm {\mathcal W}}^\star, {\bm Q}^\star)$ is also an optimal solution to problem~\eqref{eq:rank1_key_1}.
\end{Lemma}
The proof of Lemma~\ref{lemma:1} is relegated to Appendix~\ref{sec:appendix-B}.
Using  Lemma~\ref{lemma:1}, we establish the following main result:
\begin{Theorem}\label{theorem:1}
   There exists an optimal solution $({\bm {\mathcal W}}^\star, {\bm Q}^\star)$ to problem~\eqref{eq:rank1_key_1} such that ${\rm rank}({\bm {\mathcal W}}^\star) \leq 2$; i.e., ${\bm {\mathcal W}}^\star$ can be decomposed as $\bm W^\star {\bm W^\star}^H$ for some $ \bm W^\star \in \mathbb{C}^{(N-r)\times 2}$. Moreover, $({\bm W}^\star, {\bm Q}^\star)$ is a stationary solution or Karush-Kuhn-Tucker (KKT) solution to problem~\eqref{eq:main_dlinkeqv}.
\end{Theorem}
The proof of Theorem~\ref{theorem:1}  is shown in Appendix~\ref{sec:appendix-theorem1}.
The idea behind the proof is to use the semidefinite program (SDP) rank-reduction result~\cite{Huang2009} to show the existence of a rank-two optimal ${\bm {\mathcal W}}^\star$ for problem~\eqref{eq:rank1_key_1}. Theorem~\ref{theorem:1} not only pins down the SDR tightness, it also gives a procedure of recovering a stationary solution to the SSRM problem~\eqref{eq:main_dlinkeqv}. Specifically, after the convergence of the MM iteration, if $\bar{\bm{\mathcal W}}$ has rank no greater than two, we can simply obtain a stationary solution to problem~\eqref{eq:main_dlinkeqv} by applying square-root (and rank-two) decomposition to $\bar{\bm{\mathcal W}}$; otherwise, we form the problem in \eqref{eq:rank1_key_3}  and apply the rank-reduction procedure to obtain a rank-two beamforming solution to problem~\eqref{eq:main_dlinkeqv} with stationarity guarantee. We should mention that the rank-reduction procedure can be efficiently performed (with polynomial-time complexity); readers are referred to~\cite{Huang2009} for details.

To summarize, we have developed an MM approach for computing a stationary solution to the SSRM problem~\eqref{eq:main_dlinkeqv} iteratively.
However, as one may have noticed, each MM iteration in \eqref{eq:dc_sub} requires solving a convex optimization problem to optimality, which could be computationally demanding.
In view of this drawback, in the next section we will propose a low-complexity MM alternative via inexact MM updates.


\section{An Inexact MM Approach to the SSRM Problem~\eqref{eq:main_prob}} \label{sec:in-dc}

In this section, we present an inexact MM algorithm for the SSRM problem.
In addition to computational efficiency, the inexact MM algorithm to be presented is guaranteed to converge to a stationary solution to the SSRM problem.


\subsection{A General Inexact MM Framework}
Let us first introduce the notion of {\it gradient mapping}~\cite{Nesterov_book}, which will be useful for characterizing the solution inexactness and stationarity of the method to be considered.
Consider an optimization problem
\begin{equation} \label{eq:proj_grad_def}
\begin{aligned}
  \max_{\bm x} & ~ \varphi(\bm x) \\
   {\rm s.t.} & ~ \bm x \in \cal C.
  \end{aligned}
\end{equation}
where the objective function $\varphi(\bm x)$ is continuously differentiable (not necessarily convex), and the feasible set $\mathcal C$ is convex and compact.
The gradient mapping of $\varphi(\bm x)$ at $\bar{\bm x} \in \cal C$ is defined as
\begin{equation} \label{eq:proj_grad_def2}
   \tilde{\nabla} \varphi(\bar{\bm x}) \triangleq \mathcal{P}\left( \bar{\bm x} + \nabla \varphi(\bar{\bm x}) \right) - \bar{\bm x},
 \end{equation}
where ${\mathcal P}(\bm x)$ represents the projection of $\bm x$ on $\cal C$.
The following result characterizes the relationship between gradient mapping and stationarity:
\begin{Lemma}[\hspace{-.5pt}\cite{Bertsekas}]\label{lemma:stationarity}
A point $\bar {\bm x} \in \cal C$ is a stationary solution to problem~\eqref{eq:proj_grad_def}  if and only if
$ \tilde{\nabla} \varphi(\bar{\bm x}) = \bm 0$.
\end{Lemma}
Now, let us turn back to the MM subproblem~\eqref{eq:dc_sub}, which is restated below:
\vspace{-5pt}
\begin{equation}\label{eq:dc_sub_inexact}
\begin{aligned}
  \max_{\bm x} & ~\tilde{\phi}(\bm x; \bm x^{k}) \\
   {\rm s.t.} & ~ \bm x\in \mathcal D,
   \end{aligned}
\end{equation}
where, for notational convenience, we denote $\bm x \triangleq (\bm{\mathcal W}, \bm Q)$ and ${\cal D} \triangleq \{ (\bm{\mathcal W}, \bm Q)~|~{\rm Tr}(\bm{\mathcal W} + \bm Q) \leq P_R, ~\bm{\mathcal W} \succeq \bm 0, ~\bm Q \succeq \bm 0 \}$. Instead of solving problem~\eqref{eq:dc_sub_inexact}
exactly,
we perform
an update via the following rule:
\begin{center}
\fbox{
\parbox{0.46\textwidth}{
\emph{Find a point $\bm x^{k+1} \in \cal D$ such that
\begin{equation} \label{eq:inexact_key_ineq}
  \tilde{\phi}(\bm x^{k+1};\bm x^k) - \tilde{\phi}(\bm x^{k};\bm x^k) \geq \zeta^k \| \tilde{\nabla} \tilde{\phi}(\bm x^{k};\bm x^k) \|^2,
\end{equation}
where $\zeta^k>0$
is an iteration-dependent constant and is bounded away from zero.}
}
}
\end{center}
We call \eqref{eq:inexact_key_ineq} an inexact updating rule;
the reason is that the exact MM update (or the optimal solution to problem \eqref{eq:dc_sub_inexact}) satisfies \eqref{eq:inexact_key_ineq}, but the converse is not true.
We will specify in the next subsection how we build an efficient update that satisfies \eqref{eq:inexact_key_ineq}.
For now, let us focus on the guarantee of convergence to a stationary solution.
We have the following result:
\begin{Prop}\label{prop:inexact_convergence}
  Suppose that $\{\bm x^k\}$ is a sequence generated by
  an inexact updating rule in \eqref{eq:inexact_key_ineq}.
  Then, every limit point of $\{\bm x^k\}$ is a stationary solution to problem~\eqref{eq:main_eqv_sdr}.
\end{Prop}
The key of the proof of Proposition~\ref{prop:inexact_convergence} is that the updating rule~\eqref{eq:inexact_key_ineq} ensures  sufficient improvement between the consecutive iterations if the current point is not stationary.
By accumulating these improvements, the inexact MM iteration will finally reside at a stationary solution. The detailed proof is given in Appendix~\ref{sec:appendix_inexact_convergence}. In light of Proposition~\ref{prop:inexact_convergence},   a result similar to Theorem~\ref{theorem:1} is established as follows.
\begin{Theorem}\label{theorem:2}
Let $\hat{\bm x} = (\hat{\bm {\mathcal W}}, \hat{\bm Q})$ be any limit point generated by an inexact MM rule in \eqref{eq:inexact_key_ineq}, and consider the problems~\eqref{eq:rank1_key_1} and \eqref{eq:rank1_key_3} with $(\bar{\bm {\mathcal W}}, \bar{\bm Q})$ replaced by $(\hat{\bm {\mathcal W}}, \hat{\bm Q})$. Then, it holds true that there exists an optimal solution $({\bm {\mathcal W}}^\star, \bm Q^\star)$ to problem~\eqref{eq:rank1_key_1} such that ${\bm {\mathcal W}}^\star = \bm W^\star {\bm W^\star}^H$ for some $\bm W^\star \in \mathbb{C}^{(N-r)\times 2}$, and that $(\bm W^\star, {\bm Q}^\star)$ is a stationary solution to the  SSRM problem~\eqref{eq:main_dlinkeqv}.
\end{Theorem}
The proof of Theorem~\ref{theorem:2}  is identical to that of Theorem~\ref{theorem:1} and thus is omitted.

The inexact MM updating rule in \eqref{eq:inexact_key_ineq} provides a general sufficient condition under which an algorithm can guarantee convergence to a stationary solution.
The next question is how we can achieve  \eqref{eq:inexact_key_ineq} in a computationally efficient manner.
This will be addressed in the next subsection.


\subsection{A Projected Gradient-based Inexact MM Implementation}

Let us consider the following: generate an inexact solution $\bm x^{k+1}$ to problem \eqref{eq:dc_sub_inexact} via an iteration-limited projected gradient method (PGM).
As we will see shortly, such a PGM has strong connection to the aforementioned inexact MM updating rule.
The inexact PGM for problem~\eqref{eq:dc_sub_inexact} is summarized as follows:
%
%
\begin{algorithm}
  \caption{An Inexact PGM for Problem~\eqref{eq:dc_sub_inexact}}\label{algorithm:1}
\begin{algorithmic}[1]
  \State Set $l=0$, $\bm x^{k,0}=\bm x^k$ and  the maximum number of PG operations $L_k \geq 1$
   \While {$l \leq L_k-1$}
  \State Set $\bm x^{k,l+1} = \bm x^{k,l} + \alpha^{k,l} \tilde{\nabla} \tilde{\phi}(\bm x^{k,l}; \bm x^k)$, where $\alpha^{k,l}>0$ is the stepsize determined by either Armijo's rule or (limited) minimization rule~\cite{Bertsekas}.
\State $l=l+1$;
\EndWhile
  \State $\bm x^{k+1} = \bm x^{k,L_k}$.
\end{algorithmic}
\end{algorithm}

In Algorithm~\ref{algorithm:1}, the parameter $L_k$ represents the maximum number of PG operations at the $k$th MM iteration. We see that when $L_k=1$ for all $k$, Algorithm~\ref{algorithm:1} reduces to directly applying the projected gradient ascent method to the original relaxed SSRM problem~\eqref{eq:main_eqv_sdr}.
When $L_k>1$, Algorithm~\ref{algorithm:1}  has an incentive to make more progress at
each MM subproblem  by performing multiple PG operations.
Also, when every  $L_k$ approaches infinity, Algorithm~\ref{algorithm:1} becomes the exact MM update, and thus the resulting MM iteration is guaranteed to converge to a stationary solution to problem~\eqref{eq:main_dlinkeqv} by Theorem~\ref{theorem:1}.
The following proposition reveals that the aforementioned convergence guarantee holds for any finite $L_k$:
%
%
%
\begin{Prop}\label{prop:PGM_convergence}
  Suppose that $\{\bm x^k\}$ is a sequence generated by Algorithm~\ref{algorithm:1}. Then, every limit point of $\{\bm x^k\}$ is a stationary solution to problem~\eqref{eq:main_eqv_sdr}. Moreover, by using the same construction [i.e., problems~\eqref{eq:rank1_key_1} and \eqref{eq:rank1_key_3}] as that in Theorem~\ref{theorem:1}, a stationary solution to problem~\eqref{eq:main_dlinkeqv} can be extracted from every limit point of $\{\bm x^k\}$.
\end{Prop}
The key of the proof is to show that the iterations generated by Algorithm~\ref{algorithm:1} fulfill the inequality~\eqref{eq:inexact_key_ineq}. Consequently, the result follows directly from Proposition~\ref{prop:inexact_convergence} and Theorem~\ref{theorem:2}. The detailed proof is relegated to Appendix~\ref{sec:appendix_PGM_convergence}.


Thus far, we have only
considered convergence guarantees arising from Algorithm~\ref{algorithm:1}. The remaining issue is whether Algorithm~\ref{algorithm:1} can be efficiently implemented. Clearly, the main computation lies in performing the PG operations, particularly, the computation of $\tilde{\nabla} \tilde{\phi}(\bm x^{k,l};\bm x^k)$ (cf.~line 3 of Algorithm~\ref{algorithm:1}). From the definition of gradient mapping [cf.~\eqref{eq:proj_grad_def2}], one needs to find an efficient way to
calculate ${\cal P} ( \bm x^{k,l} + \nabla \tilde{\phi}(\bm x^{k,l};\bm x^k)  )$, i.e., solving the following projection problem:
\begin{equation} \label{eq:proj_cal}
\begin{aligned}
   \min_{\bm{\mathcal W} \succeq \bm 0, \bm Q \succeq \bm 0} ~ & \left\|\begin{bmatrix}
    \bm{\mathcal W}\\
    \bm Q
  \end{bmatrix}  -   \begin{bmatrix}
    \bm{\mathcal W}^{k,l} + \nabla_{\bm{\mathcal W}} \tilde{\phi}(\bm x^{k,l}; \bm  x^k)\\
    \bm Q^{k,l} + \nabla_{\bm Q} \tilde{\phi}(\bm x^{k,l}; \bm  x^k)
  \end{bmatrix} \right\|^2    \\
  {\rm s.t.}  ~ & {\rm Tr}(\bm{\mathcal W}+  \bm Q) \leq P_R,  \quad \bm{\mathcal W}\succeq \bm 0, \quad \bm Q \succeq \bm 0.
\end{aligned}
\end{equation}
Fortunately, problem~\eqref{eq:proj_cal} admits a water-filling-like solution~\cite[Fact 1]{Li_jsac}:
\begin{equation} \label{eq:wf_sol}
   \bm{\mathcal W}^\star = \bm F_1 {\rm Diag}({\bm \eta}_1^\star) \bm F_1^H, \quad \bm Q^\star =  \bm F_2 {\rm Diag}(\bm \eta_2^\star) \bm F_2^H,
 \end{equation}
where $ \bm F_1 {\rm Diag}(\tilde{\bm \eta}_1) \bm F_1^H $ and $ \bm F_2 {\rm Diag}(\tilde{\bm \eta}_2) \bm F_2^H$ are the eigenvalue decompositions of $\bm{\mathcal W}^{k,l} + \nabla_{\bm{\mathcal W}} \tilde{\phi}(\bm x^{k,l}; \bm  x^k)$ and $ \bm Q^{k,l} + \nabla_{\bm Q} \tilde{\phi}(\bm x^{k,l}; \bm  x^k)$, resp., and
\[ \bm \eta_1^\star = [\tilde{\bm \eta}_1 - \nu^\star \bm 1]^+, \quad \bm \eta_2^\star = [\tilde{\bm \eta}_2 - \nu^\star \bm 1]^+,\]
with $\nu^\star\geq 0$ being the water-filling level. The value of $\nu^\star$ relates to the total power $P_R$ and can be efficiently determined. Readers are referred to~\cite[Fact 1]{Li_jsac} for the details of solving the projection problem~\eqref{eq:proj_cal}.

\subsection{Complexity Comparison with Exact MM}
Let us analyze the computational complexities of the exact MM and the inexact MM with iteration-limited PG.
While the MM subproblem~\eqref{eq:dc_sub} is convex, it is not in a standard SDP form, owing to the logarithm function $f$.
To solve problem~\eqref{eq:dc_sub}, a successive approximation method embedded with a primal-dual interior-point method (IPM) is employed, say by \texttt{CVX}.
As is known, the arithmetic complexity for the generic  primal-dual IPM to solve a standard SDP is
${\cal O}(\max\{m, n\}^4 n^{1/2} \log(1/\varepsilon))$~\cite{SDR}, where $m$, $n$ and $\varepsilon$ represent the number of linear constraints, the dimension of the PSD cone and the solution accuracy, resp. Therefore, for the MM subproblem~\eqref{eq:dc_sub}, the
per-iteration
complexity is
${\cal O}( L_{SA} (N-r)^{4.5} \log(1/\varepsilon))$, where $L_{SA}$ denotes the number of successive approximations used. On the other hand, for the PG-based inexact MM algorithm, its computation is mainly dominated by the projection operation, which involves the eigendecomposition of complexity ${\cal O}((N-r)^3)$ and the water-filling level computation of complexity ${\cal O}(N-r)$. Therefore, the
 per-iteration
complexity of PG-based inexact MM algorithm is
 ${\cal O}( L_{PG} (N-r)^{3} )$, where $L_{PG}$ denotes the number of PG operations used for each MM subproblem. By comparing the above two complexity results, we see that the inexact MM generally has lower complexity than the exact MM, because  $L_{PG}$ is typically
  ${\cal O}(1)$, which is much smaller than $L_{SA}$.

\section{Extension: Sum Secrecy Rate Maximization under An Energy-Harvesting Eve} \label{sec:ssrm-eh}

In this section, we consider an extended scenario where the system setup and the secrecy rate model are the same as in Sec.~\ref{sec:system}, with the addition that Eve is also an energy harvesting user of the system.
Thus, the relay is also required to provide certain wireless power transfer to Eve.

Under the signal model in Sec.~\ref{sec:system}, the wireless power transfer from the source and the relay to Eve
can be formulated as
\begin{equation*}
\begin{aligned}
  & p_E(\bm W_0, \bm Q_0 ) \\
   = & \tau \Big( \frac{1}{2} \bm h_{RE}^H \mathbb{E}\{ \bm X_R(n) \bm X_R(n)^H\} \bm h_{RE} + \sum_{i\in \{A, B\}}p_i|h_{iE}|^2 \Big) \\
   = & \tau \Big( \bm h_{RE}^H (\zeta \bm W_0 \bm W_0^H + \bm Q_0 ) \bm h_{RE} + \sum_{i\in \{A, B\}}p_i|h_{iE}|^2 \Big) ,
  \end{aligned}
\end{equation*}
where $0< \tau \leq 1$ denotes the wireless power transfer efficiency; see~\cite{Krikidis}.
The subsequent SSRM problem with energy harvesting, coined SSRM-EH for short, is as follows:
\begin{subequations} \label{eq:ssrm-eh}
  \begin{align}
     \max_{\bm Q_0 \succeq \bm 0, ~\bm W_0 \in \mathbb{C}^{N\times 2}} ~~ & R_s(\bm W_0, \bm Q_0)
     \label{eq:ssrm-eh-a} \\
   {\rm s.t.} ~ ~&  p_R({\bm W}_0, {\bm Q}_0) \leq P_R,  \label{eq:ssrm-eh-b}\\
   ~ ~ & p_E(\bm W_0, \bm Q_0 ) \geq \epsilon,  \label{eq:ssrm-eh-c} \\
    ~ ~&    \bm H_{RR} \bm W_0 = \bm 0 , ~ \bm H_{RR} \bm Q_0 = \bm 0, \label{eq:ssrm-eh-d}
  \end{align}
\end{subequations}
where $\epsilon>0$ is the minimal power transfer requirement at Eve. Similar to
problem \eqref{eq:main_dlinkeqv},
the SSRM-EH problem can be equivalently written as
\begin{subequations} \label{eq:ssrm_eh_eqv}
  \begin{align}
 \hspace{-10pt}   \max_{ \bm Q \succeq \bm 0,~{\bm W} \in \mathbb{C}^{(N-r) \times 2}}&  ~   \phi(\bm W \bm W^H, \bm Q)  \label{eq:ssrm_eh_eqv_a} \\
    {\rm s.t.} & ~  {\rm Tr}(\bm W \bm W^H) + {\rm Tr}(\bm Q) \leq P_R,   \label{eq:ssrm_eh_eqv_b}\\
     ~ &  \tau  \bm h_{RE}^H \bm V_0 (  \bm W \bm W^H + \bm Q ) \bm V_0^H \bm h_{RE}   \geq \tilde{\epsilon}, \label{eq:ssrm_eh_eqv_c}
  \end{align}
\end{subequations}
where $\tilde{\epsilon} = \epsilon - \tau \sum_{i\in \{A, B\}}p_i|h_{iE}|^2 $.
Again, the SDR-based MM approach developed in Sec.~ \ref{sec:exact_dc} can be employed to handle the SSRM-EH problem~\eqref{eq:ssrm_eh_eqv}. In particular, the corresponding MM subproblem is given by
\begin{equation} \label{eq:ssrm_eh_dc_sub}
  \begin{aligned}
& (    \bm{\mathcal W}^{k+1}, \bm Q^{k+1} ) \\
 \in &   \argmax_{\bm{\mathcal W} \succeq \bm 0, \bm Q \succeq \bm 0} ~  \tilde{\phi}(\bm{\mathcal W}, \bm Q; \bm{\mathcal{W}}^k, \bm Q^k)  \\
   &~~~~~~~ {\rm s.t.}  ~~~  {\rm Tr}(\bm{\mathcal W}) + {\rm Tr}(\bm Q) \leq P_R,   \\
   & ~~~~~~~~ ~~~ ~~~\tau \bm h_{RE}^H \bm V_0 ( \bm{\mathcal W}  + \bm Q ) \bm V_0^H \bm h_{RE} \geq \tilde{\epsilon},
  \end{aligned}
\end{equation}
where $\tilde{\phi}$ is defined in \eqref{eq:dc_sub_a}.

Let $(\bar{\bm{\mathcal W}}, \bar{\bm Q})$  be  any limit point of the MM iteration in~\eqref{eq:ssrm_eh_dc_sub}. Then, a stationary solution to the SSRM-EH problem~\eqref{eq:ssrm_eh_eqv} can be retrieved from any optimal solution to the following SDP problem:
 \begin{subequations}\label{eq:rank1_key_4}
    \begin{align}
    \min_{\bm{\mathcal W} \succeq \bm 0, \bm Q \succeq \bm 0} ~ & {\rm Tr} (\bm{\mathcal W}) + {\rm Tr}(\bm Q)  \label{eq:rank1_key_4a}\\
    {\rm s.t.}~ & \eqref{eq:rank1_key_3b}-\eqref{eq:rank1_key_3e}~\text{satisfied},  \label{eq:rank1_key_4b} \\
    ~ &  \tau \bm h_{RE}^H \bm V_0 ( \bm{\mathcal W}  + \bm Q ) \bm V_0^H \bm h_{RE} \geq \tilde{\epsilon}, \label{eq:rank1_key_4c}
  \end{align}
\end{subequations}
Specifically, we have a similar result as Theorem~\ref{theorem:1}:
\begin{Theorem}\label{theorem:ssrm-eh}
 There exists an optimal solution $({\bm {\mathcal W}}^\star, {\bm Q}^\star)$
to   problem~\eqref{eq:rank1_key_4} such that ${\rm rank}({\bm {\mathcal W}}^\star) \leq 2$; i.e., ${\bm {\mathcal W}}^\star$ can be decomposed as $\bm W^\star {\bm W^\star}^H$ for some $ \bm W^\star \in \mathbb{C}^{(N-r)\times 2}$. Also, $({\bm W}^\star, {\bm Q}^\star)$ is a stationary solution or KKT solution to problem~\eqref{eq:ssrm_eh_eqv}.
\end{Theorem}
\noindent{\it Proof.}~See Appendix~\ref{appendix:theorem_ssrm_eh}. \hfill $\blacksquare$

%
%

In addition, the inexact MM update in Algorithm~\ref{algorithm:1} can  also be applied to the SSRM-EH problem~\eqref{eq:ssrm_eh_eqv} with the gradient projection step modified as
\begin{subequations} \label{eq:proj_cal_Eh}
\begin{align}
  \min_{\bm{\mathcal W} \succeq \bm 0, \bm Q \succeq \bm 0} ~ & \theta(\bm{\mathcal W} , \bm Q) \triangleq \left\|\begin{bmatrix}
    \bm{\mathcal W}\\
    \bm Q
  \end{bmatrix}  -   \begin{bmatrix}
    \bm{\mathcal W}^{k,l} + \nabla_{\bm{\mathcal W}} \tilde{\phi}(\bm x^{k,l}; \bm  x^k)\\
    \bm Q^{k,l} + \nabla_{\bm Q} \tilde{\phi}(\bm x^{k,l}; \bm  x^k)
  \end{bmatrix} \right\|^2 \label{eq:proj_cal_Eh_a} \\
  {\rm s.t.}  ~ & {\rm Tr}(\bm{\mathcal W}+  \bm Q) \leq P_R,  \label{eq:proj_cal_Eh_b}\\
  ~ & \tau \bm h_{RE}^H \bm V_0 ( \bm{\mathcal W}  + \bm Q ) \bm V_0^H \bm h_{RE} \geq \tilde{\epsilon}. \label{eq:proj_cal_Eh_c}
\end{align}
\end{subequations}
By invoking the solution in \eqref{eq:wf_sol}, problem~\eqref{eq:proj_cal_Eh} can be efficiently solved using the dual ascent method~\cite{Bertsekas}. In particular, let $\lambda\geq 0$ be the dual variable associated with~\eqref{eq:proj_cal_Eh_c}. The dual of problem~\eqref{eq:proj_cal_Eh} is given by
\[ \max_{\lambda \geq 0}  d(\lambda),\]
where
\begin{equation}\label{eq:proj_dual}
  \begin{aligned}
    & d(\lambda)   \triangleq  \\
     & \min_{\bm{\mathcal W}, \bm Q}   ~ \theta(\bm{\mathcal W} , \bm Q) - \lambda(\tau{\rm Tr}((\bm{\mathcal W}+ \bm Q )\bm V_0^H \bm h_{RE} \bm h_{RE}^H \bm V_0 ) - \tilde{\epsilon}) \\
    & ~~~{\rm s.t.}  ~ {\rm Tr}(\bm{\mathcal W}+ \bm Q)  \leq P_R, ~ \bm{\mathcal W} \succeq \bm 0,~ \bm Q \succeq \bm 0.
  \end{aligned}
\end{equation}
Given $\lambda$, problem~\eqref{eq:proj_dual} can be written into a form like problem~\eqref{eq:proj_cal}, and thus has a solution like~\eqref{eq:wf_sol}. Moreover, the optimal dual variable can be computed by using bisection to search for a $\lambda^\star\geq 0$ such that the complementarity condition for the constraint~\eqref{eq:proj_cal_Eh_c} is satisfied.

\section{Numerical Results} \label{sec:simulation}
In this section, we use Monte Carlo simulations to evaluate the performances of the proposed
SSRM algorithms.

%
%
%
%
\subsection{The Case of No Energy Harvesting with Eve}
We consider the scenario in Sec.~\ref{sec:system} and the SSRM approach in Secs.~\ref{sec:exact_dc}--\ref{sec:in-dc}.
The results to be presented in this subsection are based on
the following simulation settings, unless otherwise specified: The number of transmit antennas and receive antennas at the relay are  $N=6$ and $M=3$, resp.; all the channels are randomly generated following i.i.d. complex Gaussian distribution with zero mean  and unit variance;  the receive noise at each node has the same unit variance, i.e., $\sigma_A^2 = \sigma_B^2 = \sigma_E^2= \sigma_R^2 =1$;
both Alice and Bob have the same full-duplex SI residual factor $\kappa_A=\kappa_B=\kappa$ and the same transmit power $p_A= p_B$.


\subsubsection{Exact and Inexact MM Comparisons}
\begin{figure}[!h]
\centerline{\resizebox{.4\textwidth}{!}{\includegraphics{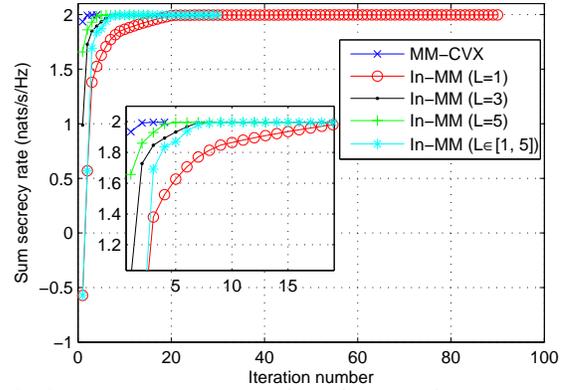}}
} \vspace*{-.5\baselineskip} \caption{Sum secrecy rate vs. iteration number ($p_A=p_B=P_R=10$dB, $\kappa=0.1$).} \label{fig:iter_rate}
\end{figure}

Fig.~\ref{fig:iter_rate} shows the convergence behaviors of the exact MM  and the inexact MM (In-MM) algorithms under one problem instance. Specifically, the exact MM solves the problem~\eqref{eq:dc_sub} exactly with \texttt{CVX}~\cite{cvx} (thus named ``MM-CVX'').
The inexact MM approximately solves the problem~\eqref{eq:dc_sub} by Algorithm~\ref{algorithm:1}, with $L_k=L$ for all $k$ and with the stepsize $\alpha^{k,l}$  determined by Armijo's rule.
Both the exact  and inexact MMs are initialized by $\bm{\mathcal  W}^0=\bm Q^0 = \frac{P_R}{2(N-r)} \bm I$. The stopping criterion for MM-CVX is $ |\tilde{\phi}(\bm x^{\bar k}; \bm x^{\bar{k}-1}) - \tilde{\phi}(\bm x^{\bar{k}-1};\bm x^{\bar{k}-2})|/|\tilde{\phi}(\bm x^{\bar{k}-1};\bm x^{\bar{k}-2})| < 10^{-3}$ for some $\bar{k}$, and
the stopping criterion for In-MM is that the In-MM iterate attains the same objective value $ \tilde{\phi}(\bm x^{\bar k}; \bm x^{\bar{k}-1})$ as the MM-CVX after convergence.
In Fig.~\ref{fig:iter_rate}, we also considered In-MMs with different number of PG operations $L$, including fixed $L=1,3,5$ and variable $L$ which is randomly and uniformly chosen from $1$ to $5$ at each MM iteration.
 As seen,
 the sum secrecy rates of both the exact and inexact MMs
 increase with the number of iterations,
 and
 converge to
 about 2 nats/s/Hz.
The exact MM converges in 4 iterations, which is very fast.
Also, the inexact MMs need more iterations to converge, varying from $7$ iterations $({\rm w.r.t.~} L=5)$ to $90$ iterations $({\rm w.r.t.~} L=1)$, which is expected.

\begin{figure}[!h]
\centerline{\resizebox{.4\textwidth}{!}{\includegraphics{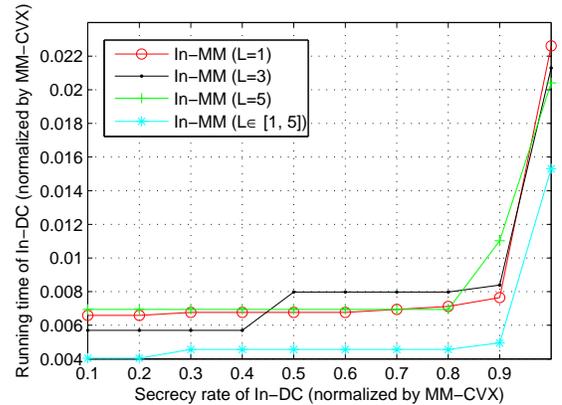}}
} \vspace*{-.5\baselineskip} \caption{Running time comparison ($p_A=p_B=P_R=10$dB, $\kappa=0.1$).} \label{fig:time}
\end{figure}

Since the per-iteration complexities of the exact MM and inexact MM
 are different, a  fairer comparison is to measure  their running times.
 Fig.~\ref{fig:time} plots the running times of In-MMs (normalized by the  time of MM-CVX at convergence) when In-MMs achieve $\alpha \tilde{\phi}(\bm x^{\bar{k}};\bm x^{\bar{k}-1})$ for $\alpha = 0.1\sim 1$ under the same setting as Fig.~\ref{fig:iter_rate}. It is clear that In-MMs run much faster than MM-CVX. Moreover, In-MM with variable $L$ is seen to be more efficient than that with fixed $L$. The reason for this is as follows: The inexact MM algorithm involves two loops,
namely, the
 outer MM iterations and the inner iteration-limited PG operations. Therefore, the total computational complexity equals the complexity of the inner PG operations times the total number of outer MM iterations. From Fig.~\ref{fig:iter_rate}, we see that the more PG operations performed for the inner loop, the less MM iterations for the outer, and vice versa.
 Therefore, there is a trade off between the solution inexactness and the number of outer MM iterations. From Fig.~\ref{fig:time}, it seems that choosing $L$ uniformly and randomly may get a better  balance of these two.


\subsubsection{Secrecy Rates Versus the Source Power}
\begin{figure}[!h]
\centerline{\resizebox{.4\textwidth}{!}{\includegraphics{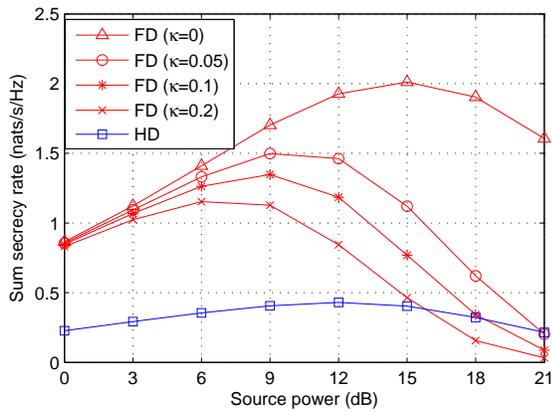}}
} \vspace*{-.5\baselineskip} \caption{Sum secrecy rate vs. the source power ($P_R=10$dB).} \label{fig:rate_source}
\end{figure}

We study the relationship between the source power and the  sum secrecy rate performance under different  SI residual level $\kappa$. For comparison, we also considered the half-duplex two-way relay designs, where Alice, Bob and the relay are all half duplex. In such a case, there is no SI at the each node, but the rate suffers from a reduction by a half. One can check that the SDR-based MM approach developed in this paper is still applicable by setting $\kappa_A =\kappa_B = 0$, removing the zero-forcing constraint~\eqref{eq:main_prob_c}, lifting the variable dimension of $\bm{\mathcal W}$ and $\bm Q$ from $(N-r)\times (N-r)$ to $N\times N$ and modifying   Eve's sum rate accordingly.\footnote{Under the half-duplex case, the received signal models at Alice, Bob and Eve are the same as before, except for the  noise covariance in~\eqref{eq:mimo_channel_Eve}, which  is changed as
\begin{align*}
  & \bm \Psi =
 \begin{bmatrix}
    \tilde{\sigma}_R^2  \| \bm h_{RE}^H \bm W_0\|^2 + \bm h_{RE}^H \bm Q_0 \bm h_{RE} + \sigma_E^2   & \\
    &    \sigma_E^2
  \end{bmatrix}.
  \end{align*}Hence, the Eve's sum rate is calculated accordingly  as

$R_{E} (\bm W_0, \bm Q_0)
  =   \log \Big(  \big(\sigma_E^2 \tilde{\sigma}_R^2+ \theta_1 \tilde{\sigma}_R^2+   \sigma_E^2 \sum_{i\in \{A,B\}} \tilde{p}_i + \theta_2 \big)\| \bm h_{RE}^H \bm W_0\|^2 +
     (\theta_1 + \sigma_E^2) \bm h_{RE}^H \bm Q_0 \bm h_{RE}  + \sigma_E^4 + \sigma_E^2 \theta_1\Big)  - \log \Big( \sigma_E^2    \tilde{\sigma}_R^2  \| \bm h_{RE}^H \bm W_0\|^2 +  \sigma_E^2 \bm h_{RE}^H \bm Q_0 \bm h_{RE} + \sigma_E^4   \Big)$.
}
Fig.~\ref{fig:rate_source}  shows the result, where ``FD'' and ``HD'' correspond to the full-duplex and half duplex-based designs, resp.
From these  figures, we have the following observations.
Firstly, we observe that the sum secrecy rate of   FD is generally better than that of HD. The reason for this is two-fold: 1) The HD protocol suffers from   half rate reduction; 2) The existence of the direct links makes the HD  more vulnerable to interception  than the FD, as the direct links of the former are free of interference, whereas for the FD case, they are interfered by the relay-to-Eve link, which somehow can better protect the sources' transmissions. Second,  the sum secrecy rates of  FD and HD both first increase with the source power, and then decrease when the source power is higher than a certain level. For the FD case, this is because the residual SI increases with the source power and can compensate any SINR gains obtained from transmit optimization; while for the HD case, this behavior is owing to the improved interception quality from the  direct  links.

\subsubsection{Secrecy Rates Versus the Number of Relay's Transmit Antenna}
\begin{figure}[!h]
\centerline{\resizebox{.4\textwidth}{!}{\includegraphics{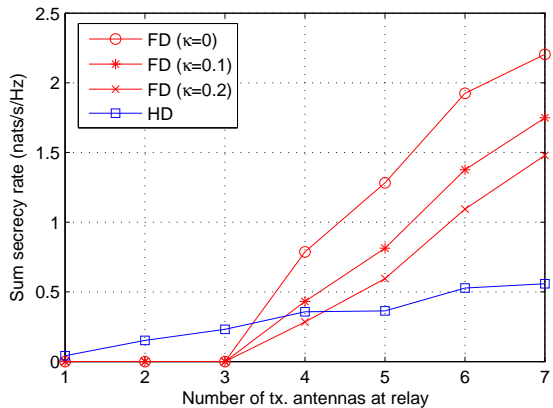}}
} \vspace*{-.5\baselineskip} \caption{Sum secrecy rate vs. the number of tx. antennas at the relay ($p_A=p_B=p_R=10$dB).} \label{fig:rate_ant}
\end{figure}

In this example, we study the relationship between the sum secrecy rate and the number of transmit antennas $N$ at the relay. The result is shown in Fig.~\ref{fig:rate_ant}. As seen, for $N\leq 3$ FD cannot provide positive secrecy rate, owing to the ZF constraints (recall the number of receive antennas at relay is 3). However, when $N$ increases, the effect of ZF constraint becomes less, and the benefit of exploiting FD (or the temporal degrees of freedom (TDoF) with STR) outweighs the loss of the spatial DoF (SDoF).

\subsection{The Energy-Harvesting Eve Case}

We consider the energy-harvesting Eve extension in Sec.~\ref{sec:ssrm-eh}.
The simulation settings are basically the same as before, i.e., $N=6$, $M=3$, $p_A = p_B$, $\kappa_A = \kappa_B = \kappa$, but with the following modifications, which are adopted to better capture the power transfer scenario: The noise variance at each node is the same and equals $-50$dBm.
All the channels follow complex Gaussian distribution with mean zero. The variance of Eve's channel is set to $-10$dB, and that of all the others' channels to $-20$dB; i.e., Eve is on average closer to the relay in order to better receive energy. The power transfer efficiency is $\tau = 10\%$.

\subsubsection{Secrecy Rates Versus the Power Transfer Threshold} Let us first study the achievable rate-power region of the proposed design for different residual SI factor $\kappa$. The result is shown in Fig.~\ref{fig:rate_energy}. For comparison, we also plotted the result of the HD design. From the figure, we see that the rate-power region shrinks when the residual SI level increases. Moreover, compared with the HD design, the FD design is able to attain larger secrecy rate when SI is well suppressed, but its maximal power transfer ability is inferior   to the HD design. This is because the FD design sacrifices the SDoF in order to better exploit the TDoF. However, for the power transfer constraint in~\eqref{eq:ssrm_eh_eqv_c},
such SDoF loss
would have impact on the maximal power transfer ability, as can be seen from \[ \tau \bm h_{RE}^H \bm V_0 ( \bm{\mathcal W}  + \bm Q ) \bm V_0^H \bm h_{RE} \leq \tau P_R \|\bm V_0^H \bm h_{RE} \|^2 < \tau P_R \| \bm h_{RE} \|^2   \]
where the first inequality is due to the total power constraint at the relay, and the last inequality follows from the fact that $\bm V_0$ is a semi-unitary matrix.

\subsubsection{The Importance of Artificial Noise}
\begin{figure}[!h]
\centerline{\resizebox{.4\textwidth}{!}{\includegraphics{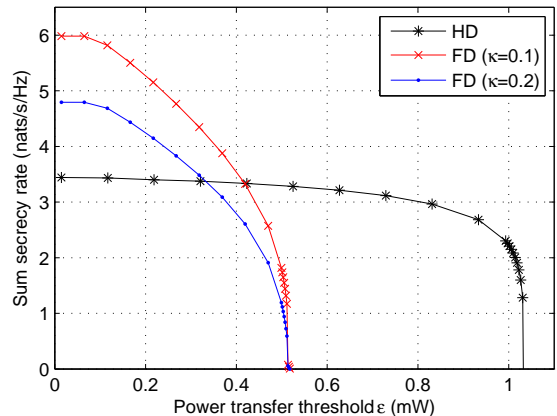}}
} \vspace*{-.5\baselineskip} \caption{Power transfer level vs. sum secrecy rate ($p_A=p_B=P_R=10$dBm).} \label{fig:rate_energy}
\end{figure}

\begin{figure}[!h]
\centerline{\resizebox{.4\textwidth}{!}{\includegraphics{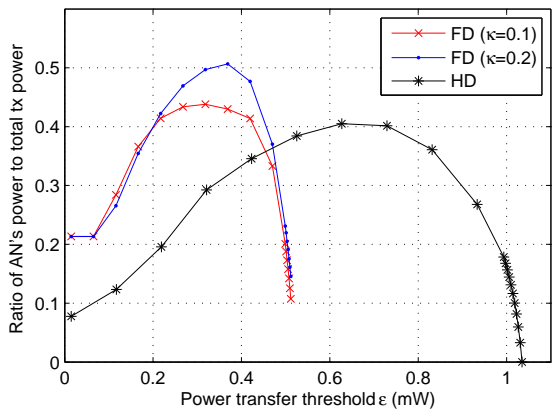}}
} \vspace*{-.5\baselineskip} \caption{The ratio of AN's power to total tx. power at relay for different power transfer threshold
$\epsilon$ ($p_A=p_B=p_R=10$dBm).} \label{fig:AN_ratio}
\end{figure}

In this example, we examine
when  AN is crucial for the design. To this end, we consider the same setting as Fig.~\ref{fig:rate_energy} and measure the ratio of AN's power to the total transmit power at the relay under different power transfer requirement $\epsilon$. The result is shown in Fig.~\ref{fig:AN_ratio}. From the figure, we see that for both FD and HD, the percentages of AN's power first increase and then decrease, when Eve's power transfer requirement increases. This phenomenon reveals an interesting result --- For extremely loose or stringent power requirements, AN is not crucial.
However, for moderate operational regions, AN is important. This may be explained as follows: For very small $\epsilon$, a small portion of AN already fulfills Eve's power transfer requirement, and there is no need to further waste power on AN. With the increase of power transfer requirement, more power needs to be allocated to Eve, and it is  reasonable to use AN to fulfill this need since it also jams Eve's reception. However, when the power transfer requirement becomes extremely stringent, the relay has to align the transmit signal around Eve (to make problem~\eqref{eq:ssrm-eh} feasible), which in turn may result in low reception power at the legitimate nodes. In other words, to Alice and Bob, the relay virtually works in a low transmission power regime. For such a power limited regime, more power should be allocated to information symbols to achieve higher secrecy rate.

\section{Conclusion}\label{sec:conclusion}

In this paper we have
considered  the sum secrecy rate maximization (SSRM)-based transmit optimization for full-duplex two-way relaying communications.
A minorization-maximization (MM) approach was proposed
for
the SSRM problem. We prove that  convergence to a stationary solution to the SSRM problem is guaranteed with either exact or inexact MM updates, as long as certain solution inexactness condition is satisfied throughout the iterations.
Extension to the case of energy-harvesting Eve was  also considered.

\section{Acknowledgement}
The authors would like to thank the anonymous reviewers
for their valuable comments which are helpful in improving
this paper.

\appendix

\subsection{Derivation of Eve's Reception Model} \label{sec:Appendix-Eve-model}

Let us focus on the reception of $s_A(2n-1)$ and $s_B(2n-1)$ from $\bm y_E(n)$ and $\bm y_E(n-1)$.
For $\bm y_E(n)$, we apply the standard Alamouti reception to retrieve a sample, denoted herein by $\tilde{y}_{E,1}(n)$, that corresponds to the reception of $s_A(2n-1)$ and $s_B(2n-1)$.
Following the model of $\bm y_E(n)$ in \eqref{eq:yEn},
one can show the expression of $\tilde{y}_{E,1}(n)$ in \eqref{eq:y_e1}, where $[\bm h_{RE}^H \bm W_0]_\ell$ denotes the $\ell$th element of $\bm h_{RE}^H \bm W_0$.
For $\bm y_E(n-1)$, observe from \eqref{eq:yEn} that $s_A(2n-1)$ and $s_B(2n-1)$ appears only in the first element of $\bm y_E(n-1)$.
By letting $\tilde{y}_{E,2}(n)= [ \bm y_E(n-1) ]_1$, and following \eqref{eq:yEn},
the expression of $\tilde{y}_{E,2}(n)$ can be obtained and it is shown in \eqref{eq:rx_sig_eve2}.
By stacking the two samples as $\tilde{\bm y}_E(n) = [ \tilde{y}_{E,1}(n) ~  \tilde{y}_{E,2}(n) ]^T$,
it can be shown from \eqref{eq:y_e1}--\eqref{eq:rx_sig_eve2} that
\[
\tilde{\bm y}(n) = \bm H_E \begin{bmatrix}
    s_A(2n-1)\\
    s_B(2n-1)
  \end{bmatrix}  + \tilde{\bm n}_E(n),
\]
where ${\bm H}_E$ and $\tilde{\bm n}_E(n)$ are defined in~\eqref{eq:mimo_channel_Eve}.

Also, the reception of $s_A(2n)$ and $s_B(2n)$ from $\bm y_E(n)$ and $\bm y_E(n-1)$ follows the same derivations as above.
Thus, we conclude that \eqref{eq:mimo_channel_Eve} provides an equivalent MIMO model for Eve's reception.

\begin{figure*}[!t]
\setlength\arraycolsep{0pt}
    \begin{equation} \label{eq:y_e1}
\begin{aligned}
\tilde{y}_{E,1}(n) & = \| \bm h_{RE}^H \bm W_0\| \left\{ \sum_{i\in \{A,B\}}    (\bm f_A+\bm f_B)^H \bm h_{iR} s_i(2n-1) +   ( \bm f_A +\bm f_B)^H n_R(2n-1) \right\} + \\
   & ~~ \frac{1}{\|\bm h_{RE}^H \bm W_0\|} \left\{ [\bm h_{RE}^H \bm W_0]_1 \Big( \bm h_{RE}^H \bm z(2n-1) + n_E(2n+1) + \sum_{i\in \{A,B\}}
   \bm h_{iE}s_i(2n+1)\Big)    -   \right. \\
    & ~~ \left. [\bm h_{RE}^H \bm W_0]_2^* \Big( \bm h_{RE}^T \bm z^*(2n) +n_E^*(2n+2) + \sum_{i\in \{A,B\}}
     \bm h_{iE}^* s_i^*(2n+2) \Big)\right\}
\end{aligned}
\end{equation}
     \hrulefill
    \end{figure*}

\begin{figure*}[!t]
\setlength\arraycolsep{0pt}
\begin{equation}\label{eq:rx_sig_eve2}
\begin{aligned}
\tilde{y}_{E,2}(n)  = & \sum_{i\in \{A,B\}}
  h_{iE} s_i(2n-1) + \bm h_{RE}^H \bm z(2n -3) + n_E(2n-1) + \\
   & [\bm h_{RE}^H \bm W_0]_1 \left\{ \sum_{i\in \{A,B\}}
   (\bm f_A+\bm f_B)^H \bm h_{iR} s_i(2n-3) +   ( \bm f_A +\bm f_B)^H \bm n_R(2n-3)\right\} + \\
    & [\bm h_{RE}^H \bm W_0]_2^* \left\{ \sum_{i\in \{A,B\}}
    (\bm f_A+\bm f_B)^H \bm h_{iR} s_i(2n-2) +   ( \bm f_A +\bm f_B)^H \bm n_R(2n-2)\right\}
  \end{aligned}
\end{equation}
     \hrulefill
    \end{figure*}

\subsection{Derivation of Eve's Sum Achievable Rate}
\label{sec:Appendix-Eve-model-2}

The sum achievable rate  \eqref{eq:rate_Eve_0} at Eve can be further derived as \eqref{eq:Re_derive},
where $\theta_1$ and $\theta_2$ are defined in~\eqref{eq:rate_Eve}.
\begin{figure*}[!t]
\setlength\arraycolsep{0pt}
\begin{equation}\label{eq:Re_derive}
\begin{aligned}
  & R_{E} (\bm W_0, \bm Q_0) \\
  = &  \log|\bm I + \bm H_E \bm P \bm H_E^H \bm \Psi^{-1}| \\
  = & \log\left|\begin{array}{cc}
    1 + \frac{ \| \bm h_{RE}^H \bm W_0\|^2(\sum_{i\in \{A, B\}} p_i |\tilde{f}_{iR}|^2 )}{\Psi_{11}} & ~~\frac{ \| \bm h_{RE}^H \bm W_0\| \sum_{i\in \{A, B\}} \tilde{f}_{iR} h_{iE}^*}{\Psi_{22}} \\
    \frac{ \| \bm h_{RE}^H \bm W_0\| \sum_{i\in \{A, B\}} \tilde{f}_{iR}^* h_{iE}}{\Psi_{11}} & ~~1 + \frac{ \sum_{i\in \{A, B\}} p_i |h_{iE}|^2}{\Psi_{22}}
  \end{array} \right|\\
  = & \log \left(  \frac{\Psi_{11}\Psi_{22} + \| \bm h_{RE}^H \bm W_0\|^2 (\sum_{i\in \{A, B\}} p_i |\tilde{f}_{iR}|^2)\Psi_{22} + \theta_1 \Psi_{11} +  \theta_2  \| \bm h_{RE}^H \bm W_0\|^2}{\Psi_{11}\Psi_{22}}\right)
\end{aligned}
\end{equation}
     \hrulefill
    \end{figure*}
To express $R_{E} (\bm W_0, \bm Q_0)$ as~\eqref{eq:rate_Eve}, notice from the definitions of $\Psi_{11} $ and $\Psi_{22}$ in~\eqref{eq:def_Psi} that $\Psi_{11} = \Psi_{22}- \| \bm h_{RE}^H \bm W_0\|^2(\tilde{p}_A+\tilde{p}_B ) + \sum_{i\in\{A,B\}} p_i|h_{iE}|^2 $. Hence,
\begin{equation}\label{eq:psi1_times_psi2}
\begin{aligned}
  & \Psi_{11}\Psi_{22}  \\
    = & \Psi_{22}^2- \| \bm h_{RE}^H \bm W_0\|^2 (\tilde{p}_A+\tilde{p}_B ) \Psi_{22} + \sum_{i\in\{A,B\}} p_i|h_{iE}|^2 \Psi_{22} \\
  = & \Psi_{22}^2- \| \bm h_{RE}^H \bm W_0\|^2 ( \textstyle \sum_{i \in \{A, B\}} p_i |\tilde{f}_{iR}|^2)\Psi_{22} + \theta_1 \Psi_{22},
\end{aligned}
\end{equation}
where the second equality is due to $\tilde{p}_i = p_i | \tilde{f}_{iR}|^2$ (cf. the definitions of $\tilde{p}_i$ and $\tilde{f}_{iR}$ for $i\in \{A,B\}$). Now, by substituting \eqref{eq:psi1_times_psi2} into \eqref{eq:Re_derive}, we obtain
\begin{align*}
  & R_{E} (\bm W_0, \bm Q_0) \\
  = & \log \left( \frac{\Psi_{22}^2 + \theta_1(\Psi_{22}+\Psi_{11}) +  \theta_2 \| \bm h_{RE}^H \bm W_0\|^2 }{\Psi_{11} \Psi_{22}} \right) .
\end{align*}

\subsection{Proof of Lemma~\ref{lemma:1}} \label{sec:appendix-B}
Since $\tilde{g}_1(\bm{\mathcal W}, \bm Q; \bm{\mathcal W}^k, \bm Q^k)$ and $\tilde{g}_2(\bm{\mathcal W}, \bm Q; \bm{\mathcal W}^k, \bm Q^k)$
 are convex upper bounds of $g_1(\bm{\mathcal W}, \bm Q)$ and $g_2(\bm{\mathcal W}, \bm Q)$, resp., which is tight at $(\bm{\mathcal W}^k, \bm Q^k)$,  it follows from the basic MM property that
  \begin{align*}
    \phi(\bm{\mathcal W}^{k}, \bm Q^{k}) &    =   \tilde{\phi}(\bm{\mathcal W}^{k}, \bm Q^{k}; \bm{\mathcal{W}}^k, \bm Q^k)  \\
     & \leq \tilde{\phi}(\bm{\mathcal W}^{k+1}, \bm Q^{k+1}; \bm{\mathcal{W}}^k, \bm Q^k) \\
     &  \leq \phi(\bm{\mathcal W}^{k+1}, \bm Q^{k+1}) \\
     & ~~\vdots \\
     & \leq \phi(\bar{\bm{\mathcal W}}, \bar{\bm Q}),
   \end{align*}
  i.e., $\{\tilde{\phi}(\bm{\mathcal W}^{k+1}, \bm Q^{k+1}; \bm{\mathcal{W}}^k, \bm Q^k)\}_k$ converges monotonically to the upper bound $\phi(\bar{\bm{\mathcal W}}, \bar{\bm Q})$. By noting the continuity of the function $\tilde{\phi}$,
and taking
a convergent subsequence of $\{(\bm{\mathcal{W}}^k, \bm Q^k)\}_k$ with a limit point $(\bar{\bm{\mathcal{W}}}, \bar{\bm Q}) $, we have
\[ \tilde{\phi}(\bm{\mathcal W}, \bm Q;\bar{\bm{\mathcal{W}}}, \bar{\bm Q}) \leq \phi(\bar{\bm{\mathcal W}}, \bar{\bm Q}) \]
for all feasible $(\bm{\mathcal W}, \bm Q)$. In addition, from the definition of $\tilde{\phi}$, we have $\tilde{\phi}(\bar{\bm{\mathcal{W}}}, \bar{\bm Q};\bar{\bm{\mathcal{W}}}, \bar{\bm Q}) =\phi(\bar{\bm{\mathcal W}}, \bar{\bm Q})$. Thus, $(\bar{\bm {\mathcal W}}, \bar{\bm Q})$ is an optimal solution to problem~\eqref{eq:rank1_key_1}.

On the other hand from the construction of problem~\eqref{eq:rank1_key_3}, one can easily show that every optimal $(\bm {\mathcal W}^\star, \bm Q^\star)$ of \eqref{eq:rank1_key_3} must be feasible for problem~\eqref{eq:rank1_key_1} with $\tilde{\phi}( {\bm{\mathcal{W}}}^\star, {\bm Q}^\star;\bar{\bm{\mathcal{W}}}, \bar{\bm Q}) = \tilde{\phi}(\bar{\bm{\mathcal{W}}}, \bar{\bm Q};\bar{\bm{\mathcal{W}}}, \bar{\bm Q})  $. Hence, every optimal $(\bm {\mathcal W}^\star, \bm Q^\star)$ of \eqref{eq:rank1_key_3} is optimal for \eqref{eq:rank1_key_1}.

\subsection{Proof of Theorem~\ref{theorem:1}} \label{sec:appendix-theorem1}
Notice that problem~\eqref{eq:rank1_key_3} is an SDP with 7 constraints.
By the SDP rank-reduction result in \cite[Lemma~3.1]{Huang2009},
there exists an optimal solution $(\bm {\mathcal W}^\star, \bm Q^\star)$ to problem~\eqref{eq:rank1_key_3} such that
\[ {\rm rank}(\bm {\mathcal W}^\star)^2 +  {\rm rank}(\bm { Q}^\star)^2 \leq 7,\]
which implies
 that
\[ {\rm rank}(\bm {\mathcal W}^\star) \leq \sqrt{7} <3 \]
Hence, ${\rm rank}(\bm {\mathcal W}^\star)$ equals 1 or 2, and by Lemma~\ref{lemma:1} such an optimal $(\bm {\mathcal W}^\star, \bm Q^\star)$ is also optimal for problem~\eqref{eq:rank1_key_1}. Next, we establish the second part of the Theorem.

Since $({\bm {\mathcal W}}^\star, \bm Q^\star) $  is an optimal solution to problem~\eqref{eq:rank1_key_1}, it satisfies the KKT conditions of problem~\eqref{eq:rank1_key_1}. To describe it, let us denote $\gamma \geq 0$, $\bm{\mathcal Y} \succeq \bm 0$ and $\bm Z  \succeq \bm 0$ as the dual variables associated with the power constraint, ${\bm{\mathcal W}}^\star \succeq \bm 0$ and ${\bm Q}^\star\succeq \bm 0 $ of problem~\eqref{eq:rank1_key_1}. Then, the KKT conditions of problem~\eqref{eq:rank1_key_1} are  shown in~\eqref{eq:proof_kkt_rank1_key1}, where for notational simplicity we have dropped all the arguments and used ${(\cdot)}^\star$ [resp. $\bar{(\cdot)}$]  to represent the function value or gradient evaluation at the point $(\bm {\mathcal W}^\star, \bm Q^\star)$ [resp. $(\bar{\bm {\mathcal W}}, \bar{\bm Q})$].
\begin{figure*}[!t]
\setlength\arraycolsep{0pt}
\begin{subequations} \label{eq:proof_kkt_rank1_key1}
\begin{align}
&    \sum_{i=1}^2 \frac{\nabla_{\bm {\mathcal W}} \alpha_i^\star}{c_i+\alpha_i^\star}
 + \sum_{i=1}^2 \frac{\nabla_{\bm {\mathcal W}}\psi_i^\star }{\psi_i^\star }
 - \sum_{i=1}^2   \frac{\nabla_{\bm{\mathcal W}}\bar{\beta}_i}{c_i+ \bar{\beta}_i}  -  \frac{  2 \psi_{2}^\star \nabla_{\bm{\mathcal W}}\psi_{2}^\star   + \theta_1( \nabla_{\bm{\mathcal W}} \psi_{1}^\star + \nabla_{\bm{\mathcal W}} \psi_{2}^\star) +  \theta_2 \zeta^{-1} \bm V_0^H  \bm h_{RE}  \bm h_{RE}^H \bm V_0}{ (  \bar{\psi}_{2} )^2 + \theta_1(  \bar{\psi}_{1} +  \bar{\psi}_{2}) +  \theta_2 \zeta^{-1}  \bm h_{RE}^H \bm V_0 \bar{\bm{\mathcal W}} \bm V_0^H \bm h_{RE} }  - \gamma \bm I + \bm{\mathcal Y} = \bm 0, \label{eq:proof_kkt_rank1_key1a} \\
 &  \sum_{i=1}^2 \frac{\nabla_{\bm {Q}} \alpha_i^\star}{c_i+\alpha_i^\star}
 + \sum_{i=1}^2 \frac{\nabla_{\bm {Q}}\psi_i^\star }{\psi_i^\star }
 - \sum_{i=1}^2   \frac{\nabla_{\bm{Q}}\bar{\beta}_i}{c_i+ \bar{\beta}_i}  -  \frac{  2 \psi_{2}^\star \nabla_{\bm Q}\psi_{2}^\star   + \theta_1( \nabla_{\bm Q} \psi_{1}^\star + \nabla_{\bm Q} \psi_{2}^\star) }{ (  \bar{\psi}_{2} )^2 + \theta_1(  \bar{\psi}_{1} +  \bar{\psi}_{2}) +  \theta_2 \zeta^{-1}  \bm h_{RE}^H \bm V_0 \bar{\bm{\mathcal W}}\bm V_0^H \bm h_{RE} } - \gamma \bm I + \bm Z = \bm 0, \label{eq:proof_kkt_rank1_key1b}\\
& {\rm Tr}( {\bm{\mathcal W}}^\star + {\bm Q}^\star) \leq P_R,  \quad {\bm Q}^\star \succeq \bm 0, \quad \gamma \geq 0, \label{eq:proof_kkt_rank1_key1c} \\
& {\bm Q}^\star \bm Z = \bm 0, \quad \bm Z \succeq \bm 0, \label{eq:proof_kkt_rank1_key1d}\\
&  \bm {\mathcal Y}  {\bm{\mathcal W}}^\star  = \bm 0, \quad  \bm{\mathcal Y} \succeq \bm 0,  \quad  {\bm{\mathcal W}}^\star \succeq \bm 0. \label{eq:proof_kkt_rank1_key1e}
\end{align}
\end{subequations}
     \hrulefill
    \end{figure*}

Since  $\beta_i^\star  = \bar{\beta}_i$ and $\psi_i^\star  = \bar{\psi}_i$ for all $i$  [cf. problem~\eqref{eq:rank1_key_3}], and   $\nabla_{\bm {\mathcal W}} \bar{\beta}_i$,  $\nabla_{\bm Q} \bar{\beta}_i$ are  constant matrices, irrespective of $({\bm{\mathcal W}}, {\bm Q})$, Eq.~\eqref{eq:proof_kkt_rank1_key1a} and \eqref{eq:proof_kkt_rank1_key1b} can be reexpressed as~\eqref{eq:proof_kkt_rank1a} and \eqref{eq:proof_kkt_rank1b}, resp.
\begin{figure*}[!t]
\setlength\arraycolsep{0pt}
\begin{subequations} \label{eq:proof_kkt_rank1}
\begin{align}
&    \sum_{i=1}^2 \frac{\nabla_{\bm {\mathcal W}} \alpha_i^\star}{c_i+\alpha_i^\star}
 + \sum_{i=1}^2 \frac{\nabla_{\bm {\mathcal W}}\psi_i^\star }{\psi_i^\star }
 - \sum_{i=1}^2   \frac{\nabla_{\bm{\mathcal W}} {\beta}_i^\star}{c_i+ {\beta}_i^\star}  -  \frac{  2 \psi_{2}^\star \nabla_{\bm{\mathcal W}}\psi_{2}^\star   + \theta_1( \nabla_{\bm{\mathcal W}} \psi_{1}^\star + \nabla_{\bm{\mathcal W}} \psi_{2}^\star) +  \theta_2 \zeta^{-1}  \bm V_0^H \bm h_{RE}  \bm h_{RE}^H \bm V_0}{ (   {\psi}_{2}^\star )^2 + \theta_1(  {\psi}_{1}^\star +  {\psi}_{2}^\star) +  \theta_2 \zeta^{-1}  \bm h_{RE}^H \bm V_0  {\bm{\mathcal W}}^\star \bm V_0^H \bm h_{RE} } - \gamma \bm I + \bm{\mathcal Y} = \bm 0,    \label{eq:proof_kkt_rank1a} \\
&  \sum_{i=1}^2 \frac{\nabla_{\bm {Q}} \alpha_i^\star}{c_i+\alpha_i^\star}
 + \sum_{i=1}^2 \frac{\nabla_{\bm {Q}}\psi_i^\star }{\psi_i^\star }
 - \sum_{i=1}^2   \frac{\nabla_{\bm{Q}} {\beta}_i^\star}{c_i+  {\beta}_i^\star}  -  \frac{  2 \psi_{2}^\star \nabla_{\bm Q}\psi_{2}^\star   + \theta_1( \nabla_{\bm Q} \psi_{1}^\star + \nabla_{\bm Q} \psi_{2}^\star) }{  (   {\psi}_{2}^\star )^2 + \theta_1(  {\psi}_{1}^\star +  {\psi}_{2}^\star) +  \theta_2 \zeta^{-1}  \bm h_{RE}^H \bm V_0  {\bm{\mathcal W}}^\star \bm V_0^H \bm h_{RE} } - \gamma \bm I + \bm Z = \bm 0.   \label{eq:proof_kkt_rank1b}
\end{align}
\end{subequations}
     \hrulefill
    \end{figure*}
Moreover, from the previous proof, we have shown that ${\rm rank}({\bm {\mathcal W}}^\star) \leq 2$. Thus, ${\bm {\mathcal W}}^\star$ can be decomposed as ${\bm {\mathcal W}}^\star = \bm W^\star {\bm W^\star}^H$ for some $\bm W^\star \in \mathbb{C}^{(N-r)\times 2}$. It thus follows from~\eqref{eq:proof_kkt_rank1_key1e} that
\begin{equation} \label{eq:proof_kkt_key_complement}
  \bm {\mathcal Y}  \bm W^\star  {\bm W^\star}^H   = \bm 0  \Longleftrightarrow \bm {\mathcal Y}  \bm W^\star  = \bm 0.
\end{equation}
By postmultiplying the both sides of \eqref{eq:proof_kkt_rank1a} with $2\bm W^\star$, and using  \eqref{eq:proof_kkt_key_complement}, we arrive at~\eqref{eq:proof_kkt_1st_order}.
\begin{figure*}[!t]
\setlength\arraycolsep{0pt}
\begin{equation} \label{eq:proof_kkt_1st_order}
\begin{aligned}
 & \sum_{i=1}^2 \frac{2 \nabla_{\bm {\mathcal W}} \alpha_i^\star \bm W^\star}{c_i+\alpha_i^\star}
 + \sum_{i=1}^2 \frac{2 \nabla_{\bm {\mathcal W}}\psi_i^\star  \bm W^\star }{\psi_i^\star }
 - \sum_{i=1}^2   \frac{2 \nabla_{\bm{\mathcal W}} {\beta}_i^\star  \bm W^\star}{c_i+ {\beta}_i^\star}  \\
 & -  \frac{  4 \psi_{2}^\star \nabla_{\bm{\mathcal W}}\psi_{2}^\star  \bm W^\star   + 2\theta_1( \nabla_{\bm{\mathcal W}} \psi_{1}^\star + \nabla_{\bm{\mathcal W}} \psi_{2}^\star)  \bm W^\star +  2 \theta_2 \zeta^{-1}  \bm V_0^H \bm h_{RE}  \bm h_{RE}^H \bm V_0  \bm W^\star}{ (   {\psi}_{2}^\star )^2 + \theta_1(  {\psi}_{1}^\star +  {\psi}_{2}^\star) +  \theta_2 \zeta^{-1}  \bm h_{RE}^H \bm V_0   \bm W^\star {\bm W^\star}^H \bm V_0^H \bm h_{RE} }    = 2\gamma \bm W^\star
   \end{aligned}
\end{equation}
     \hrulefill
    \end{figure*}
Moreover, one can verify that the following equations hold:
\begin{equation}\label{eq:proof_kkt_diff_equality}
\begin{aligned}
   & 2\nabla_{\bm {\mathcal W}} \alpha_i (\bm {\mathcal W}^\star, \bm {Q}^\star)  \bm W^\star = \nabla_{\bm W} \alpha_i(\bm W^\star {\bm W^\star}^H, \bm Q^\star)~i=1,2,\\
   & 2\nabla_{\bm {\mathcal W}} \beta_i (\bm {\mathcal W}^\star, \bm {Q}^\star)  \bm W^\star = \nabla_{\bm W} \beta_i(\bm W^\star {\bm W^\star}^H, \bm Q^\star)~i=1,2,\\
   &   2 \nabla_{\bm{\mathcal W}}\psi_{i}(\bm {\mathcal W}^\star, \bm {Q}^\star)  \bm W^\star =   \nabla_{\bm{  W}}\psi_{i}(\bm W^\star {\bm W^\star}^H, \bm Q^\star), ~i=1,2.
 \end{aligned}
 \end{equation}
By substituting~\eqref{eq:proof_kkt_diff_equality} into \eqref{eq:proof_kkt_1st_order},
and by replacing $\bm{\mathcal W}^\star$ with $\bm W^\star {\bm W^\star}^H$ in \eqref{eq:proof_kkt_rank1_key1c} and \eqref{eq:proof_kkt_rank1b},
 we see that \eqref{eq:proof_kkt_rank1_key1c}, \eqref{eq:proof_kkt_rank1_key1d}, \eqref{eq:proof_kkt_rank1b} and \eqref{eq:proof_kkt_1st_order} exactly
 constitute the KKT conditions of problem~\eqref{eq:main_dlinkeqv}.
 Therefore, $(\bm W^\star, \bm Q^\star)$, together with the dual variables $\gamma, \bm Z$, forms a KKT point of problem~\eqref{eq:main_dlinkeqv}.

\subsection{Proof of Proposition~\ref{prop:inexact_convergence}} \label{sec:appendix_inexact_convergence}
Recall $\phi(\bm x^{k+1}) \geq  \tilde{\phi}(\bm x^{k+1};\bm x^k)$ and    $\phi(\bm x^k) = \tilde{\phi}(\bm x^{k};\bm x^k)$. We  have
\begin{align*}
   \phi(\bm x^{k+1}) - \phi(\bm x^{k}) & \geq \tilde{\phi}(\bm x^{k+1};\bm x^k) - \tilde{\phi}(\bm x^{k};\bm x^k) \\
    & \geq \zeta^k \| \tilde{\nabla} \tilde{\phi}(\bm x^{k};\bm x^k) \|^2 \\
    & \geq 0,
 \end{align*}
i.e., $\{ \phi(\bm x^{k} )\}$ is nondecreasing.
Taking summation on both sides of the above inequality over $k$ from $0$ to $K-1$  yields
\[ \phi(\bm x^K) - \phi(\bm x^0) \geq \sum_{k=0}^{K-1} \zeta^k \|\tilde{\nabla} \tilde{\phi}(\bm x^{k};\bm x^k) \|^2.\]
Since $\cal D$ is compact and $\phi$ is continuous, $ \phi(\bm x^K) - \phi(\bm x^0)$ is  bounded from above. Moreover, because $\zeta^k$ is bounded away from zero as $k\rightarrow \infty$, it must hold that
\begin{equation} \label{eq:proof_inexact_dc}
  \lim_{k\rightarrow \infty} \tilde{\nabla} \tilde{\phi}(\bm x^{k};\bm x^k) = \bm 0.
\end{equation}
Due to the compactness of $\cal D$,  the sequence $\{\bm x^k\}$ has at least one limit point, say  $\bar{\bm x}$.
Moreover, because $\tilde{\phi}$ is continuously differentiable and the projection operation is a continuous mapping~\cite[Prop. B.11(c)]{Bertsekas}, their composite $\tilde{\nabla}\tilde{\phi}$ is a continuous mapping, which together with \eqref{eq:proof_inexact_dc} implies
\[ \tilde{\nabla} \tilde{\phi}(\bar{\bm x}; \bar{\bm x}) = \bm 0. \]
On the other hand, notice that
\begin{align*}
  \tilde{\nabla} \tilde{\phi}(\bar{\bm x}; \bar{\bm x}) & = \mathcal{P}\left( \bar{\bm x} + \nabla \tilde{\phi}(\bar{\bm x}) \right) - \bar{\bm x} \\
  & =  \mathcal{P}\left( \bar{\bm x} + \nabla {\phi}(\bar{\bm x}) \right) - \bar{\bm x} \\
  & = \tilde{\nabla} {\phi}(\bar{\bm x}; \bar{\bm x}),
\end{align*}
where the second equality is due to the fact that $\tilde{\phi}$ is a partial linearization  of ${\phi}$. Therefore, we obtain $ \tilde{\nabla} {\phi}(\bar{\bm x}; \bar{\bm x}) = \bm 0$; i.e., $\bar{\bm x}$ is a stationary point to problem~\eqref{eq:main_eqv_sdr}.

\subsection{Proof of Proposition~\ref{prop:PGM_convergence}} \label{sec:appendix_PGM_convergence}
For ease of exposition, we prove only for the real variable case; extension to the  complex domain is straightforward.
Let us first consider the Armijo's stepsize rule; i.e., $\alpha^{k,l} = (\beta_{k,l})^{m_{k,l}}$
for some constant $\beta_{k,l} \in (0,~1)$, where the integer $m_{k,l}$ is chosen as the smallest nonnegative integer such that the following inequality holds~\cite{Bertsekas}
\begin{equation}\label{eq:proof_prop2}
  \begin{aligned}
  & \tilde{\phi}\left(\bm x^{k,l}+ (\beta_{k,l})^{m_{k,l}} \tilde{\nabla} \tilde{\phi}(\bm x^{k,l};\bm x^k) ;\bm x^k \right) - \tilde{\phi}(\bm x^{k,l};\bm x^k)  \\
  \geq & \sigma (\beta_{k,l})^{m_{k,l}} \left( \nabla \tilde{\phi}(\bm x^{k,l};\bm x^k)\right)^T \tilde{\nabla} \tilde{\phi}(\bm x^{k,l};\bm x^k),
\end{aligned}
\end{equation}
for some constant $\sigma\in (0,~1)$. Next, we bound the right-hand side of \eqref{eq:proof_prop2} by using the following  lemma~\cite[Prop.~2.1.3]{Bertsekas}:
\begin{Lemma}[Projection Theorem] \label{lemma:VI}
Let $\cal X$ be a nonempty, closed and convex subset of $\mathbb{R}^N$. Given some $\bm x \in \mathbb{R}^N$ and its projection $\bar{\bm x}$ onto $\cal X$, i.e., $\bar{\bm x} = {\cal P}(\bm x)$, it holds that
\[ (\bm x - \bar{\bm x})^T (\bm z - \bar{\bm x}) \leq 0, \quad \forall ~\bm z \in \cal X. \]
\end{Lemma}
Now, by substituting $\bm x =\bm x^{k,l} + \nabla\tilde{\phi}(\bm x^{k,l}; \bm x^k)$ and $\bm z = \bm x^{k,l}$ in Lemma~\ref{lemma:VI} and denoting $\bar{\bm x}^{k,l} = {\cal P} \left(\bm x^{k,l} + \nabla\tilde{\phi}(\bm x^{k,l}; \bm x^k) \right)$,  we have
\[ (   \bm x^{k,l} + \nabla\tilde{\phi}(\bm x^{k,l}; \bm x^k ) -  \bar{\bm x}^{k,l} )^T ( \bm x^{k,l} - \bar{\bm x}^{k,l}) \leq 0. \]
Rearranging the above inequality yields
\[
\nabla\tilde{\phi}(\bm x^{k,l}; \bm x^k)^T (\bar{\bm x}^{k,l} - \bm x^{k,l})  \geq \| \bar{\bm x}^{k,l} -\bm x^{k,l} \|^2,    \]
i.e.,
\begin{equation}\label{eq:vi}
\nabla\tilde{\phi}(\bm x^{k,l}; \bm x^k)^T \tilde{\nabla}\tilde{\phi}(\bm x^{k,l}; \bm x^k)  \geq \| \tilde{\nabla}\tilde{\phi}(\bm x^{k,l}; \bm x^k)\|^2.
\end{equation}
Combining Eqn.~\eqref{eq:proof_prop2} and \eqref{eq:vi}, we obtain
\begin{align*}
& \tilde{\phi}\left(\bm x^{k,l}+ (\beta_{k,l})^{m_{k,l}} \tilde{\nabla} \tilde{\phi}(\bm x^{k,l};\bm x^k) ;\bm x^k \right) - \tilde{\phi}(\bm x^{k,l};\bm x^k) \\
\geq &  \sigma (\beta_{k,l})^{m_{k,l}} \| \tilde{\nabla}\tilde{\phi}(\bm x^{k,l}; \bm x^k)\|^2.
\end{align*}
Recalling $\bm x^{k,l+1} = \bm x^{k,l} + (\beta_{k,l})^{m_{k,l}} \tilde{\nabla} \tilde{\phi}(\bm x^{k,l};\bm x^k)$ and taking summation  over $l$ from $0$ to $L_k-1$ yields
\begin{align*}
  \tilde{\phi}(\bm x^{k+1};\bm x^k) - \tilde{\phi}(\bm x^{k};\bm x^k) & \geq \sigma \sum_{l=0}^{L_k-1} (\beta_{k,l})^{m_{k,l}} \| \tilde{\nabla}\tilde{\phi}(\bm x^{k,l}; \bm x^k)\|^2 \\
  & \geq  \underbrace{\sigma  (\beta_{k,0})^{m_{k,0}}}_{\triangleq \zeta^k} \| \tilde{\nabla}\tilde{\phi}(\bm x^{k}; \bm x^k)\|^2.
\end{align*}
Assuming for  now that $\sigma  (\beta_{k,0})^{m_{k,0}}$ is bounded away from zero as $k\rightarrow \infty$ (we will prove this shortly),  we see that using Armijo's stepsize rule, the iteration fulfills the relation~\eqref{eq:inexact_key_ineq}. Therefore, the convergence of Algorithm~\ref{algorithm:1} follows directly from Proposition~\ref{prop:inexact_convergence}.

Now, we show that $\sigma  (\beta_{k,0})^{m_{k,0}}$ is indeed bounded away from zero as $k\rightarrow \infty$. The proof is inspired by Proposition 1.2.1 in~\cite{Bertsekas}.
Consider a convergent subsequence of $\{ \bm x^k\}$, denoted by $\{ \bm x^k\}_{k \in \cal K}$, with a limit point $\bar{\bm x}$, i.e., $\lim_{k\rightarrow \infty, k\in {\cal K}} \bm x^k = \bar{\bm x}$. Suppose on the contrary that $\limsup_{k\rightarrow \infty, k\in \cal K}\sigma  (\beta_{k,0})^{m_{k,0}} = 0$, i.e.,
\[ \limsup_{k\rightarrow \infty, k\in \cal K}    (\beta_{k,0})^{m_{k,0}} = 0. \]
Hence, by the definition of Armijo's rule, we must have for some index $\bar{k}$ such that
\begin{equation} \label{eq:proof_armijo}
\begin{aligned}
& \tilde{\phi}\left(\bm x^{k,0}+ \frac{(\beta_{k,0})^{m_{k,0}}}{\beta_{k,0}} \tilde{\nabla} \tilde{\phi}(\bm x^{k,0};\bm x^k) ;\bm x^k \right) - \tilde{\phi}(\bm x^{k,0};\bm x^k) \\
<&  \sigma \frac{(\beta_{k,0})^{m_{k,0}}}{\beta_{k,0}} \left( \nabla \tilde{\phi}(\bm x^{k,0};\bm x^k)\right)^T \tilde{\nabla} \tilde{\phi}(\bm x^{k,0};\bm x^k),
\end{aligned}
   \end{equation}
for all $k\in {\cal K}, ~k \geq \bar{k}$. Denote $\bm p^k = \tilde{\nabla} \tilde{\phi}(\bm x^{k,0};\bm x^k)/ \| \tilde{\nabla} \tilde{\phi}(\bm x^{k,0};\bm x^k) \|$ and $\bar{\alpha}^{k,0} =  (\beta_{k,0})^{m_{k,0}}  \| \tilde{\nabla} \tilde{\phi}(\bm x^{k,0};\bm x^k) \| /\beta_{k,0}$. Since $\tilde{\phi}(\bm x; \bm x^k)$ is continuously differentiable and the feasible set $\cal D$ is compact [cf.~Eqn.~\eqref{eq:dc_sub_inexact}], $\| \tilde{\nabla} \tilde{\phi}(\bm x^{k,0};\bm x^k) \|$ is bounded, and thus
\[\limsup_{k\rightarrow \infty, k\in \cal K}\bar{\alpha}^{k,0} = 0.\]
In addition, since $\| \bm p^k \| =1$, it has a limit point. By taking a further subsequence of $\cal K$, we can assume without loss of generality that $\lim_{k\rightarrow \infty, k\in \cal K} \bm p^k = \bar{\bm p}$. Now, by substituting $\bm p^k$ and $\bar{\alpha}^{k,0}$ into \eqref{eq:proof_armijo}, we have
\begin{align*}
   \frac{\tilde{\phi}\left(\bm x^{k,0}+ \bar{\alpha}^{k,0} \bm p^k ;\bm x^k \right) - \tilde{\phi}(\bm x^{k,0};\bm x^k)}{\bar{\alpha}^{k,0}} <  & \sigma  {\nabla} \tilde{\phi}(\bm x^{k,0};\bm x^k)^T \bm p^k, \\
   & ~~~~  \forall~k\in {\cal K}, ~k\geq \bar{k},
 \end{align*}
which further implies
\begin{align*}
   \nabla \tilde{\phi}\left(\bm x^{k,0} + \tilde{\alpha}^{k,0} \bm p^k; \bm x^k \right)^T \bm p^k < & \sigma   {\nabla} \tilde{\phi}(\bm x^{k,0};\bm x^k)^T\bm p^k, \\
   &~~~~~~~~~ \forall~k\in {\cal K}, ~k\geq \bar{k},
 \end{align*}
for some $\tilde{\alpha}^{k,0} \in [0,~\bar{\alpha}^{k,0}]$ by applying the mean value theorem. Taking the limit and noticing $\bar{\alpha}^{k,0} \rightarrow 0$ and $\bm x^{k,0} = \bm x^k$, we obtain
\[ (1-\sigma) \nabla \tilde{\phi}\left(\bar{\bm x}; \bar{\bm x} \right)^T \bar{\bm p} \leq 0,  \]
i.e.,
\begin{equation} \label{eq:proof_armijo_key}
  \nabla \tilde{\phi}\left(\bar{\bm x}; \bar{\bm x} \right)^T \bar{\bm p} \leq 0 .
\end{equation}
On the other hand, it follows from \eqref{eq:vi} that
\[\nabla\tilde{\phi}(\bm x^{k,l}; \bm x^k)^T \frac{\tilde{\nabla}\tilde{\phi}(\bm x^{k,l}; \bm x^k)}{\|\tilde{\nabla}\tilde{\phi}(\bm x^{k,l}; \bm x^k) \|}  \geq \| \tilde{\nabla}\tilde{\phi}(\bm x^{k,l}; \bm x^k)\|.  \]
By setting $l=0$ and taking limit over $k\in \cal K$, we get
\[ \nabla\tilde{\phi}(\bar{\bm x}; \bar{\bm x})^T \bar{\bm p}  \geq \| \tilde{\nabla}\tilde{\phi}(\bar{\bm x}; \bar{\bm x})\|.  \]
Let us consider two possibilities for the limit point $\bar{\bm x}$: 1) if $\bar{\bm x}$ is a stationary point, then there is nothing to prove and Proposition~\ref{prop:PGM_convergence} holds trivially; 2) if $\bar{\bm x}$ is not a stationary point, then we must have
\[ \| \tilde{\nabla}\tilde{\phi}(\bar{\bm x}; \bar{\bm x})\| >0. \]
Hence, $\nabla\tilde{\phi}(\bar{\bm x}; \bar{\bm x})^T \bar{\bm p}>0$, but this contradicts with \eqref{eq:proof_armijo_key}. Therefore, $\sigma  (\beta_{k,0})^{m_{k,0}}$ is bounded away from zero.

For the (limited) minimization stepsize rule, the proof is essentially the same as that of Armijo's stepsize rule if one notice that
\begin{align*}
& \tilde{\phi}\left(\bm x^{k,l}+ \eta^{k,l} \tilde{\nabla} \tilde{\phi}(\bm x^{k,l};\bm x^k) ;\bm x^k \right) - \tilde{\phi}(\bm x^{k,l};\bm x^k) \\
\geq  & \tilde{\phi}\left(\bm x^{k,l}+ (\beta_{k,l})^{m_{k,l}} \tilde{\nabla} \tilde{\phi}(\bm x^{k,l};\bm x^k) ;\bm x^k \right) - \tilde{\phi}(\bm x^{k,l};\bm x^k)\\
\geq & \sigma (\beta_{k,l})^{m_{k,l}} \left( \nabla \tilde{\phi}(\bm x^{k,l};\bm x^k)\right)^T \tilde{\nabla} \tilde{\phi}(\bm x^{k,l};\bm x^k),
  \end{align*}
where $\eta^{k,l}>0$ is the stepsize obtained from (limited) minimization rule. Consequently, the proof for (limited) minimization rule boils down to that of Armijo's stepsize rule.

\subsection{Proof of Theorem~\ref{theorem:ssrm-eh}}\label{appendix:theorem_ssrm_eh}
The proof  is basically the same as that of Theorem~\ref{theorem:1}. Here, we just give a sketched proof. It is easy to verify that Lemma~\ref{lemma:1} still holds if we add the power transfer constraint~\eqref{eq:rank1_key_4c} in problems~\eqref{eq:rank1_key_1} and \eqref{eq:rank1_key_3}. Moreover, notice that there are in total 8 constraints in~\eqref{eq:rank1_key_4}; hence it follows from the rank-reduction result~\cite[Lemma~3.1]{Huang2009} that there exists an optimal $({\bm {\mathcal W}}^\star, {\bm Q}^\star)$ for problem~\eqref{eq:rank1_key_4} fulfilling
\[ {\rm rank}({\bm {\mathcal W}}^\star) \leq \sqrt{8} < 3.\]
That is, ${\rm rank}({\bm {\mathcal W}}^\star) \leq 2$. The remaining proof is exactly the same as that of Theorem~\ref{theorem:1} and thus omitted.

\end{document}